\documentclass{article}
 \pdfoutput=1
\usepackage{amssymb,graphicx,cancel}
\usepackage[margin=2cm]{geometry}
\usepackage{slashed}
\usepackage{hepunits}
\usepackage{color}
\usepackage[table]{xcolor}
\usepackage{graphicx}
\usepackage[normalem]{ulem}
\usepackage{jheppub}
\newcommand{\ie}{{\it i.e.~}}  
\newcommand{\eg}{{\it e.g.~}}

\newcommand{\Fig}[1]{Fig.~\ref{#1}}
\newcommand{\Eq}[1]{Eq.~\ref{#1}}
\newcommand{\Sec}[1]{Section~\ref{#1}}
\newcommand{\Tab}[1]{Table~\ref{#1}}

\preprint{IFIC/18-21, FTUV-18-0524.0283}
\title{The Dispirited Case of Gauged $U(1)_{B-L}$ Dark Matter}
\author[a]{Miguel Escudero,}
\author[a]{Samuel J. Witte}
\author[a]{and Nuria Rius}
\affiliation[a]{Departamento de F\'isica Te\'orica and IFIC, Universidad de Valencia-CSIC,
C/ Catedr\'atico Jos\'e Beltr\'an, 2, E-46980 Paterna, Spain}
\abstract{
We explore the constraints and phenomenology of possibly the simplest scenario that could account at the same time for the active neutrino masses and the dark matter in the Universe within a gauged $U(1)_{B-L}$ symmetry, namely right-handed neutrino dark matter. We find that null searches from lepton and hadron colliders require dark matter with a mass below 900 GeV to annihilate through a resonance. Additionally, the very strong constraints from  high-energy dilepton searches fully exclude the model for $ 150 \, \text{GeV} < m_{Z'} < 3 \, \text{TeV}$. We further explore the phenomenology in the high mass region (\ie masses $\gtrsim \mathcal{O}(1) \, \text{TeV}$) and highlight theoretical arguments, related to the appearance of a Landau pole or an instability of the scalar potential, disfavoring large portions of this parameter space. Collectively, these considerations illustrate that a minimal extension of the Standard Model via a local $U(1)_{B-L}$ symmetry with a viable thermal dark matter candidate is difficult to achieve without fine-tuning. We conclude by discussing possible extensions of the model that relieve tension with collider constraints by reducing the gauge coupling required to produce the correct relic abundance. 
}
\emailAdd{Miguel.Escudero@ific.uv.es}
\emailAdd{Sam.Witte@ific.uv.es}
\emailAdd{Nuria.Rius@ific.uv.es}

\keywords{}

\begin{document}
\maketitle
\flushbottom

%%%%%%%%%%%%%%%%%%%%%%%%%%%%%%%%%%%%%%%%%%%%%%%%%%%%%%%%%%%%%%%%%%%%%%%%%%%
\section{Introduction}\label{sec:Introduction}
%%%%%%%%%%%%%%%%%%%%%%%%%%%%%%%%%%%%%%%%%%%%%%%%%%%%%%%%%%%%%%%%%%%%%%%%%%%

The nature of the dark matter in the Universe and the origin of the tiny neutrino masses are two unsolved fundamental questions in particle physics. 
Small neutrino masses are naturally generated by extending the Standard Model (SM) with massive Majorana neutrinos, singlets under the SM gauge group, via the well-known 
seesaw mechanism. While the scale of such sterile neutrino masses is largely unconstrained (it can be the GUT scale for Yukawa couplings $Y$ of ${\cal O}(1)$, a few GeV for $Y \sim 10^{-6}$, etc.), the mixing among active and sterile neutrinos is generically very small, making it difficult to detect sterile neutrinos at current colliders, even if kinematically accessible. Interestingly, this simple extension of the SM can also provide a dark matter candidate if the lightest of the sterile neutrinos has a keV-scale mass \cite{Dodelson:1993je,Barbieri:1989ti,Kainulainen:1990ds,Adhikari:2016bei}. Although a very appealing possibility, this extension involving only the addition of right-handed neutrinos is disfavored by the absence of an $X$-ray signal arising from the sterile neutrino decay $N \rightarrow \nu \gamma$~\cite{Boyarsky:2007ge,Yuksel:2007xh,Perez:2016tcq} -- a process necessarily induced by the required mixing with the active neutrinos~\cite{Asaka:2005an}.

A slightly enlarged scenario that keeps the link between sterile neutrinos and dark matter consists of charging both under $U(1)_{B-L}$  and coupling them to a SM singlet scalar, also charged under this global $U(1)_{B-L}$  symmetry, which generates Majorana masses for all the SM singlet fermions. The spontaneous breaking of the $U(1)_{B-L}$ leads to the appearance of a Goldstone boson, the Majoron~\cite{Chikashige:1980ui,Schechter:1981cv}.
 In this scenario there are two potential dark matter candidates:
1) the Majoron, assuming it acquires a mass e.g. via non-perturbative gravitational effects which break the global 
 symmetry~\cite{Coleman:1988tj,Akhmedov:1992hi} (see \cite{Lattanzi:2014mia} for a recent review on the subject, and for a fresh perspective on global symmetries 
 see~\cite{Witten:2017hdv}),
 and 
2) one (or several) of the massive SM singlet fermions. 
As before, the simplest solution involves taking the dark matter to be a keV sterile neutrino. However, unless the Majoron is more massive than the sterile neutrino, the sterile neutrino is unstable\footnote{Alternatively, one can explicitly break the symmetry using a real scalar field, in which case there is no Goldstone boson associated with the spontaneous breaking of the global symmetry \cite{Petraki:2007gq}.}.
Thanks to the interactions with the scalar singlet, a heavier sterile neutrino stabilized by an additional symmetry (\eg a $Z_2$ symmetry which forbids a Yukawa coupling to SM leptons) can also play the role of the dark matter. This possibility has been 
extensively studied by two of the authors in \cite{Escudero:2016tzx}, where it was shown that the observed dark matter relic abundance can be obtained by the annihilation freeze-out of a Majorana fermion with mass in the range of a WIMP ($\sim$1 GeV - 2 TeV). Within this scenario, the common origin of the dark matter mass and the masses of the other sterile neutrinos (\ie those that mix with the active neutrinos) naturally links the unknown seesaw scale to the scale of a thermal WIMP.
 
Despite containing an electroweak scale dark matter candidate, these scenarios are often quite elusive, a consequence of the fact that the production mechanism in terrestrial experiments (other than the possible mixing of active and sterile neutrinos) arises from the mixing between the SM Higgs and the CP-even component of the singlet scalar. Given current limits on the aforementioned mixing from \eg precision Higgs measurements at the LHC (see \eg\cite{Belanger:2013xza,Brehmer:2015rna,Beniwal:2015sdl,deBlas:2016ojx,Khachatryan:2016vau}), significant production of the sterile neutrinos at colliders is unattainable. Furthermore, the presence of the massless (or light) Majoron allows for invisible decays of both the sterile neutrinos and the CP-even scalar, implying that even should these particles be produced, their dominant decay modes make this model extremely hard to test.

 As a consequence, considerably more attention has been given to models in which the SM is extended by a $U(1)_{B-L}$ gauge symmetry, 
 which is interesting from both a phenomenological perspective (in the sense that it is easier to probe experimentally) and  a theoretical point of view, since canceling the anomaly arising from thew new gauge symmetry naturally requires the existence of three right-handed neutrinos. The absence of the Majoron makes a keV sterile neutrino again cosmologically stable, and its new gauge interaction allows for the production of the observed dark matter relic 
abundance by the freeze-in mechanism, without conflicting with the non-observation of an $X$-ray signal from its decay \cite{Kaneta:2016vkq}. Alternatively, one can consider a more massive sterile neutrino dark matter candidate, once again stabilized by an additional symmetry, whose relic abundance is now determined via the more conventional freeze-out mechanism. Historically, studies focused on this scenario have only studied a small part of the viable parameter space, analyzing just one particular mediator and/or mass range (\eg the $Z'$ portal has been explored in \cite{Kaneta:2016vkq,Okada:2016gsh,Klasen:2016qux,Okada:2016tci,Okada:2018ktp}, the Higgs portal in \cite{Okada:2010wd,Basak:2013cga}, and the classically conformal minimal model in \cite{Okada:2012sg,Oda:2017kwl}).

We find that a thorough analysis of the minimal gauged $U(1)_{B-L}$ sterile neutrino\footnote{Here we use term `sterile' to refer to the right-handed neutrinos, even though they may have interactions with the Standard Model through the $B-L$ gauge interaction.} WIMP dark matter model is missing from the literature, and thus we present here a comprehensive study of its viability. We focus explicitly on the case where the dark matter fermion has a lepton charge $L=1$, but comment in the text on alternative possibilities~\cite{Montero:2007cd,Ma:2014qra,Kanemura:2014rpa,Singirala:2017see,Wang:2015saa,Wang:2017mcy}.  
To be concrete we choose a simple $Z_2$ symmetry to stabilize the dark matter candidate because, although not theoretically appealing, it is representative of more
involved models and captures the main phenomenology of Majorana dark matter  within  the context of a gauged $U(1)_{B-L}$ extension of the SM. 
 Moreover, we also study generic phenomenological features appearing in extensions of the minimal scenario via the addition of singlet scalars.

The paper is organized as follows. In \Sec{sec:Model} we briefly review the minimal $U(1)_{B-L}$  dark matter framework considered here.
After summarizing in \Sec{sec:constraints} all current constraints on the $U(1)_{B-L}$ gauge boson $Z^\prime$, 
we focus in \Sec{sec:DMpheno} on the dark matter phenomenology and  identify, via a comprehensive scan, the viable parameter space within the minimal scenario for which a dark matter candidate can still be accommodated. In \Sec{sec:ThC} 
we summarize the results of the scan and discuss various theoretical concerns regarding the allowed parameter space, \eg the potential appearance of a Landau pole or induced instability of the scalar sector occurring at low energies. 
\Sec{sec:ext} is devoted to the analysis of generic extensions of the minimal framework. We conclude in \Sec{sec:Conclusions}.

%%%%%%%%%%%%%%%%%%%%%%%%%%%%%%%%%%%%%%%%%%%%%%%%%%%%%%%%%%%%%%%%%%%%%%%%%%%
\section{Model Outline}\label{sec:Model}
%%%%%%%%%%%%%%%%%%%%%%%%%%%%%%%%%%%%%%%%%%%%%%%%%%%%%%%%%%%%%%%%%%%%%%%%%%%
In this work we consider possibly the simplest gauged $U(1)_{B-L}$ extension of the Standard Model with sterile neutrinos and a dark matter candidate. We construct the model based on the minimal particle content required for $U(1)_{B-L}$ to be non-anomalous. Namely, we introduce three right handed leptons with charge $L = 1$. Two ($N_1,N_2$) of these three states will be the ones responsible for the masses and oscillatory pattern of the active neutrinos via the type-I seesaw mechanism. The other ($\chi$) will be our dark matter particle, which is stabilized via a $Z_2$ symmetry. In order to give masses to these fermions we also introduce a scalar field ($\phi$) with charge $L = -2$, which upon symmetry breaking will render Majorana masses for these right-handed leptons. The relevant part of the Lagrangian then reads:
\begin{eqnarray}\label{eq:Lagrangian}
{\cal L} &\supset& \mu_H^2 H^\dagger H - \lambda_H (H^\dagger H)^2  + 
\mu_\phi^2 \phi^\dagger \phi - \lambda_\phi (\phi^\dagger \phi)^2 - 
\lambda_{H\phi} (H^\dagger H) \, (\phi^\dagger \phi)   \\
&-& \left( \frac{\lambda_{\chi}}{\sqrt{2}} \phi \, \overline{\chi}_{R} \chi_{R}^c + h.c. \right) - 
\left( \frac{\lambda_{N_1}}{\sqrt{2}} \phi \, \overline{N}_{R1} N_{R1}^c + h.c. \right) - 
\left( \frac{\lambda_{N_2}}{\sqrt{2}} \phi \, \overline{N}_{R2} N_{R2}^c + h.c. \right) \nonumber \\
&-& (Y_{\alpha 1} \overline {L}_L^\alpha H N_{R1}  + h.c.)- (Y_{\alpha 2} \overline {L}_L^\alpha H N_{R2}  + h.c.) \nonumber \, .
\end{eqnarray} 
Additionally, it should be understood that the inclusion of the new gauge symmetry necessitates a modification to the covariant derivatives which subsequently produce interactions with the new $Z^\prime$. 

Dirac dark matter candidates within the context of this (or a similar) model have been studied in \eg\cite{Duerr:2015wfa,Alves:2015mua,Klasen:2016qux,DeRomeri:2017oxa,FileviezPerez:2018toq,Han:2018zcn}. We choose to focus on a Majorana candidate because this maximally suppresses constraints from direct dark matter detection; collider constraints, on the other hand, are largely insensitive to the difference between Dirac and Majorana dark matter, making our analysis here maximally conservative. We also note that one could consider a slight non-minimal extension of this model with a scalar dark matter candidate~\cite{Rodejohann:2015lca,Klasen:2016qux,Bandyopadhyay:2017bgh,Biswas:2017tce}; however, in addition to being non-minimal, this model suffers from strong  direct detection constraints similar to the case of Dirac dark matter.

%%%%%%%%%%%%%%%%%%%%%%%%%%%%%%%%%%%%%%%%%%%%%%%%%%%%%%%%%%%%%%%%%%%%%%%%%%%
\section{Constraints on the $Z^\prime$}\label{sec:constraints}
\begin{figure}[t]
\centering
\begin{tabular}{cc}
\hspace{-0.5cm} \includegraphics[width=0.5\textwidth]{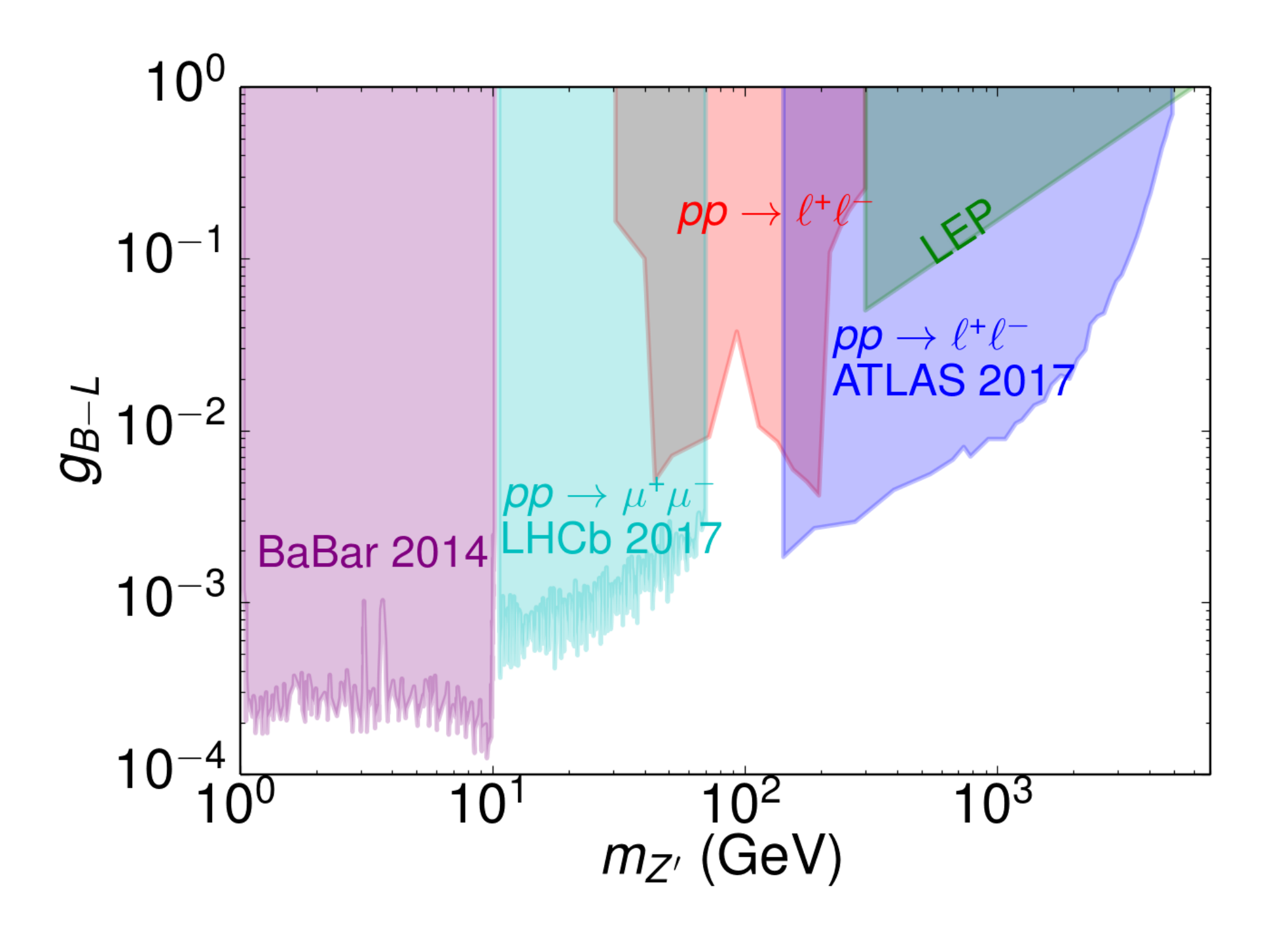} &  \includegraphics[width=0.5\textwidth]{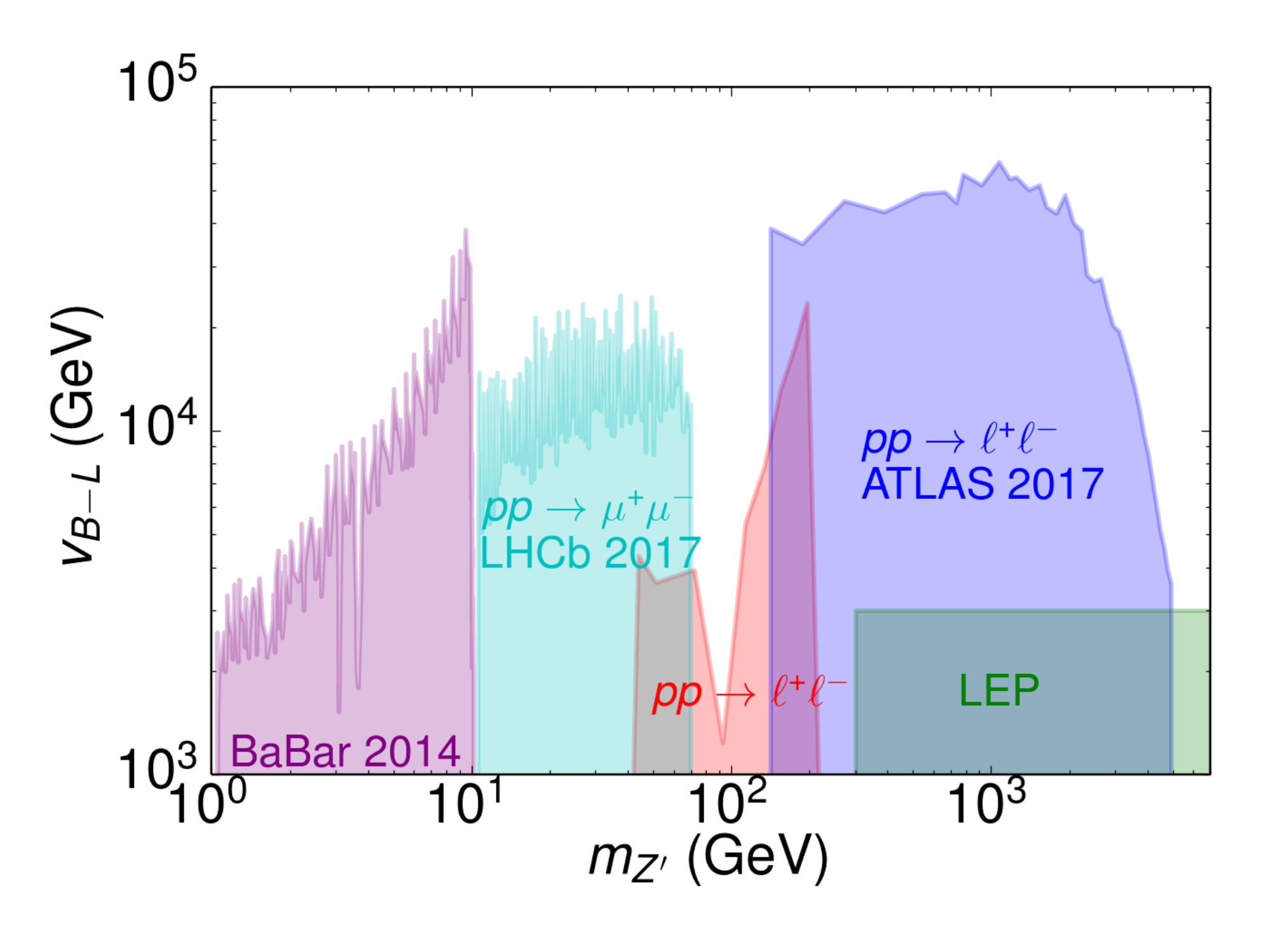}
 \end{tabular}
\caption{Bounds from various collider experiments on the ${B-L}$ $Z'$ gauge boson in terms of the coupling $g_{B-L}$ (left) and the vev $v_{B-L}$ (right).}\label{fig:Zonly}
\end{figure}

%%%%%%%%%%%%%%%%%%%%%%%%%%%%%%%%%%%%%%%%%%%%%%%%%%%%%%%%%%%%%%%%%%%%%%%%%%%
In the following subsections we enumerate the constraints that arise from having a $Z^\prime$ with non-negligible couplings to both baryons and leptons. We emphasize that the collider constraints described below implicitly assume that the invisible branching fraction of the $Z^\prime$ is negligible, however we have explicitly checked that for any choice of $m_\chi$ and $m_{N_1,N_2}$ this invisible channel does not exceed $\sim 25\%$, and thus the constraints are never significantly altered. Nevertheless, we verify in the following sections the extent to which reducing collider bounds, such that invisible decays are saturated, ameliorates existing tension on the model\footnote{Note that processes involving an off-shell $Z'$ are insensitive to the invisible branching fraction, provided the coupling strength remains fixed.}. Note that the maximally attainable invisible branching fraction of the $Z^\prime$ can be reduced should the dark matter have lepton number $L < 1$ (thus making the presented results conservative), and enhanced should $L > 1$, however for reasonable values of $L$ this enhancement is only expected to be significant near resonance.
A summary of the collider constraints on the $B-L$ gauge coupling as a function of the $Z^\prime$ mass is presented in~\Fig{fig:Zonly}.

%%%%%%%%%%%%%%%%%%%%%%%%%%%%%%%%%%%%%%%%%%%%%%%%%%%%%%%%%%%%%%%%%%%%%%%%%%%
\subsection{LHC}\label{subsec:LHC}
%%%%%%%%%%%%%%%%%%%%%%%%%%%%%%%%%%%%%%%%%%%%%%%%%%%%%%%%%%%%%%%%%%%%%%%%%%%

A number of searches from the LHC produce strong constraints on $Z^\prime$ models, in particular for models in which the $Z^\prime$ couples to charged leptons. We briefly review the most stringent constraints on our model below.

In the high mass region, \ie $150$ GeV $\lesssim m_{Z^\prime} \lesssim 4$ TeV, bounds from ATLAS on the dilepton final state using $36 {\rm fb}^{-1}$ of data at $13$ TeV lead to strong limits on the $B-L$ coupling (shown in blue in \Fig{fig:Zonly})~\cite{Aaboud:2017buh} (see~\cite{Sirunyan:2018exx} for similar CMS constraints on this channel). Constraints on the $B-L$ model are produced by the collaboration, and thus the presented bounds are inferred using Fig.~5 of~\cite{Aaboud:2017buh}. 

Constraints on lower $Z^\prime$ masses (30 GeV $\lesssim m_{Z^\prime} \lesssim $ 300 GeV) can be derived by analyzing the modification that would be produced to the Drell-Yan (DY) differential cross section. The triple differential cross section for DY production has been measured to remarkably high accuracy; here, we focus on the modifications to the invariant mass spectrum with muon final states (in the central rapidity channel), using measurements from ATLAS at $\sqrt{s} = 8$ TeV with $20.2 {\rm fb}^{-1}$~\cite{Aaboud:2017ffb}. Note that a similar analysis was performed in the context of a different model in~\cite{Bishara:2017pje}. Explicitly, we use {\tt MadGraph5 aMC@NLO}~\cite{Alwall:2014hca,Hirschi:2015iia} to compute the modified DY differential cross section for invariant dimuon masses 
$46 \,\text{GeV} < m_{\mu\mu} < 200$ GeV. Cuts on the pseudo-rapidity and transverse momentum of the $Z^\prime$ are imposed using {\tt MadAnalysis5}~\cite{Conte:2012fm}, with 10 equally spaced bins in the dimuon spectrum and assuming uncorrelated errors between bins, following the analysis of the collaboration~\cite{Aaboud:2017ffb}. The derived constraint is shown in \Fig{fig:Zonly} in red.

At yet lower masses (10.6 GeV $< m_{Z^\prime} < $ 70 GeV), recent constraints from LHCb on the identical channel, \ie $pp\rightarrow \mu^+\mu^-$, provide the dominant constraint on the $Z^\prime$~\cite{Aaij:2017rft}. Re-deriving such constraints using {\tt MadGraph5 aMC@NLO} is non-trivial, as the analysis performed by the collaboration is complicated. However, since the results are interpreted in terms of a dark photon model, such bounds can be effectively translated into the $B-L$ model, as is done \eg in~\cite{Ilten:2018crw}. This can be accomplished by noting that the production cross section from $u\bar{u}$ and $c\bar{c}$ are suppressed by a factor of 4 relative to that of the dark photon model, and production cross sections from the $d\bar{d}$, $s\bar{s}$, and $b\bar{b}$ channels are identical in both models~\cite{Ilten:2018crw}. Since the branching fractions to muons in both models are approximately equal within the mass range of interest, conservative limits can be set on the $B-L$ coupling by dividing the cross section by a factor of 4 (note that this is conservative because only some, not all, of the production channels are actually suppressed). The bounds shown in cyan in \Fig{fig:Zonly} accurately reproduce those derived in~\cite{Ilten:2018crw}.

The above list certainly does not account for the entirety of the constraints on this model; less stringent constraints from other channels are not shown here to maintain clarity (see~\cite{Batell:2016zod} for other current and future constraints). 

It is also worth noting that in generic models with multiple $U(1)$ gauge symmetries, mixing between the gauge bosons may occur. Such a mixing typically leads to even stronger constraints on the $B-L$ gauge coupling, see \eg\cite{Bandyopadhyay:2018cwu} for derived limits on the mixing in a $U(1)_{B-L}$ model and~\cite{Curtin:2014cca} for general light $Z'$ constraints from kinetic mixing. Throughout this paper we have implicitly taken this mixing to be zero, however in principle the running of the couplings may induce non-zero mixing at scales probed by colliders. We have verified that for our purposes, the evolution of this coupling for the parameter space of interest is small over the relevant scales.

Before continuing we would like to briefly comment on the potential reach of the future high-luminosity LHC (HL-LHC) in constraining the existence of such a $Z^\prime$. From the discussion above, it should be clear that dilepton searches provide the most stringent constraints; this is simply a consequence of the fact that the branching ratio to leptons is large and electron/muon searches are clean. In order to understand the extent to which the high mass region of parameter space may be probed (within the authors' lifetimes), we preform the following simple first-order analysis: using {\tt MadGraph5 aMC@NLO}, we determine the distribution of dimuon events as a function of invariant dimuon mass arising from the $Z^\prime$, taking the $B-L$ coupling to be at the perturbative limit (\ie $g_{B-L} = \sqrt{4\pi}$), setting the center-of-mass energy $\sqrt{s} = 14$ TeV, and assuming a luminosity of $\mathcal{L} = 3 \, ab^{-1}$. We find that roughly $\lesssim \mathcal{O}(5)$ events can be observed for $m_{Z^\prime} \gtrsim 7$ TeV; thus we will consider $m_{Z^\prime} \gtrsim 7$ TeV to be untestable in near-future colliders.

%%%%%%%%%%%%%%%%%%%%%%%%%%%%%%%%%%%%%%%%%%%%%%%%%%%%%%%%%%%%%%%%%%%%%%%%%%%
\subsection{BaBar}
%%%%%%%%%%%%%%%%%%%%%%%%%%%%%%%%%%%%%%%%%%%%%%%%%%%%%%%%%%%%%%%%%%%%%%%%%%%

The BaBar collaboration has published strong constraints on both visible decays~\cite{Lees:2014xha}, with $e$ and $\mu$ final states, and invisible decays~\cite{Lees:2017lec} of a dark photon in the mass range $\sim 2 m_\mu - 10$ GeV. Since the $Z^\prime$ in the $U(1)_{B-L}$ model decays predominantly to visible particles, we use this search~\cite{Lees:2014xha} to set bounds. The published limits on the visible decay modes can be straightforwardly translated into constraints on the $B-L$ gauge coupling by equating the dark photon mixing with the gauge coupling, and multiplying the constraint by the relative branching fractions in each model (for each decay channel). Since the branching fractions of the dark photon to charged leptons are approximately equal to those of the $Z^\prime$, the limits are nearly identical in strength. Furthermore, limits from BaBar on invisible decays are of comparable strength, implying bounds on $g_{B-L}$ cannot differ by more than a factor of $2$, even should the $Z^\prime$ preferentially decay to invisible particles.

%%%%%%%%%%%%%%%%%%%%%%%%%%%%%%%%%%%%%%%%%%%%%%%%%%%%%%%%%%%%%%%%%%%%%%%%%%%
\subsection{LEP}
%%%%%%%%%%%%%%%%%%%%%%%%%%%%%%%%%%%%%%%%%%%%%%%%%%%%%%%%%%%%%%%%%%%%%%%%%%%

The bound from LEP originates from the agreement of the measured $e^+ e^-\to \ell^+ \ell^-$ cross sections at various energies with the SM prediction~\cite{ALEPH:2004aa}. In particular, for the $B-L$ gauge boson, it was shown in~\cite{Appelquist:2002mw,Carena:2004xs} that LEP observations require $(m_{Z'} /\text{TeV} ) > 6 \,g_{B-L} $.

%%%%%%%%%%%%%%%%%%%%%%%%%%%%%%%%%%%%%%%%%%%%%%%%%%%%%%%%%%%%%%%%%%%%%%%%%%%
\section{Dark matter phenomenology}\label{sec:DMpheno}
%%%%%%%%%%%%%%%%%%%%%%%%%%%%%%%%%%%%%%%%%%%%%%%%%%%%%%%%%%%%%%%%%%%%%%%%%%%
In this section we comment on the relevant phenomenology of our Majorana dark matter candidate. In particular we discuss the computation of the relic density (\Sec{subsec:DMrelic}), direct detection constraints (\Sec{subsec:DD}), the algorithm used to determine the viable parameter space (\Sec{subsec:DMscan}), and the relevant annihilation phenomenology (\Sec{subsec:annhil}).

%%%%%%%%%%%%%%%%%%%%%%%%%%%%%%%%%%%%%%%%%%%%%%%%%%%%%%%%%%%%%%%%%%%%%%%%%%%
\subsection{Dark Matter Relic Density}\label{subsec:DMrelic}
\begin{figure}[t]
\centering
 \includegraphics[width=1\textwidth]{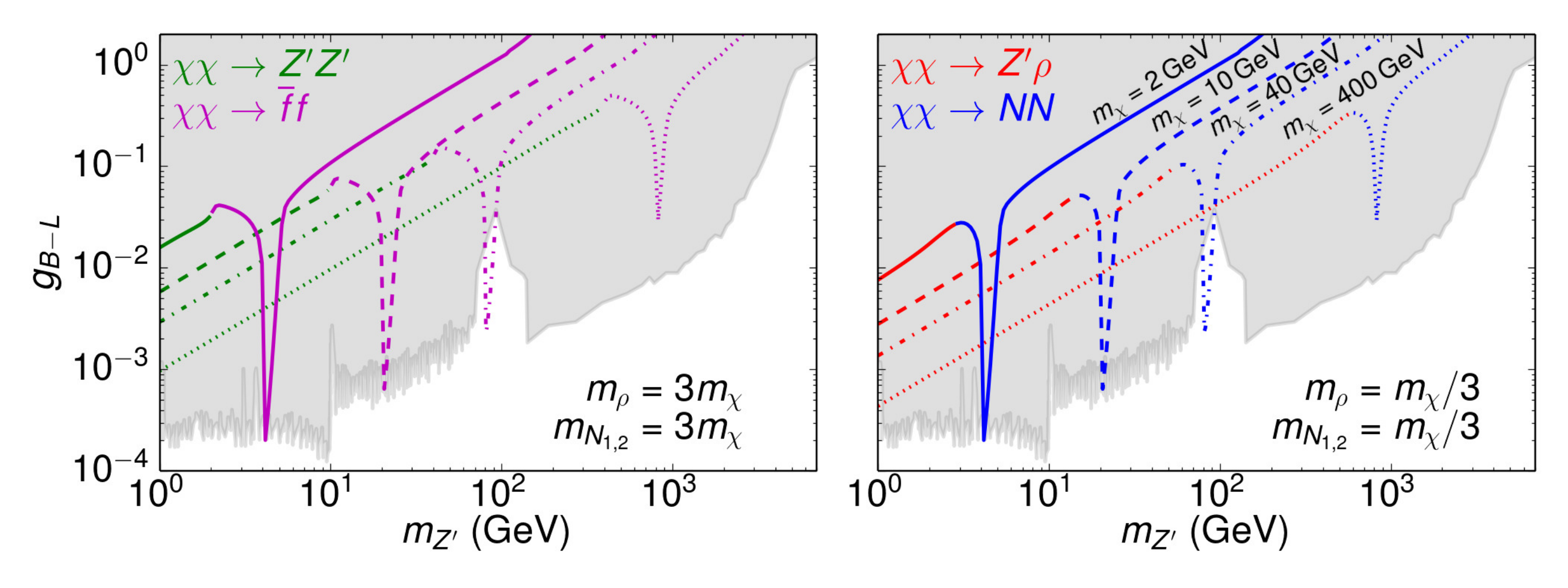} 
\caption{Coupling as a function of $Z^\prime$ mass producing $\Omega_{\chi} = 0.12$ for various choices of the dark matter mass. The shaded region identifies couplings that are excluded from $Z^\prime$ searches at colliders (see Figure~\ref{fig:Zonly}). The colors depict the dominant annihilation channel for dark matter velocities $v = 10^{-3}$, as is relevant for indirect searches. \textit{Left:} Masses are chosen such that both sterile neutrinos and the CP-even scalar are kinematically inaccessible (\ie $m_N = m_\rho= 3m_\chi$). \textit{Right:} Masses are chosen such that both sterile neutrinos and the CP-even scalar are kinematically accessible, and light with respect to the dark matter (\ie $m_N = m_\rho = m_\chi/3$). The lines in the left panel correspond to the same dark matter masses as in the right panel.}\label{fig:DMrelic}
\end{figure}

%%%%%%%%%%%%%%%%%%%%%%%%%%%%%%%%%%%%%%%%%%%%%%%%%%%%%%%%%%%%%%%%%%%%%%%%%%%
The dark matter relic density has been computed by implementing the model in \texttt{FeynRules}~\cite{Christensen:2008py} and using  \texttt{micrOMEGAs\_4.3.5}~\cite{Barducci:2016pcb}. Since the parameter space is quite large, it is not possible to show the diverse range of possible couplings that give rise to the correct dark matter abundance. For illustrative purposes, we show in Figure~\ref{fig:DMrelic} the value of the gauge coupling that produces the correct relic abundance as a function of the $Z^\prime$ mass for various choices of the dark matter, scalar and right-handed neutrino masses, plotted against the constraints discussed in \Sec{sec:constraints}. It is worth pointing out that the values of the gauge coupling producing the correct relic abundance are fairly independent of the masses of the CP-even scalar and the sterile neutrinos, provided they are not on-resonance with $\chi$. Thus, the examples provided in \Fig{fig:DMrelic} are broadly representative of the model.

%%%%%%%%%%%%%%%%%%%%%%%%%%%%%%%%%%%%%%%%%%%%%%%%%%%%%%%%%%%%%%%%%%%%%%%%%%%
\subsection{Direct Detection}\label{subsec:DD}
%%%%%%%%%%%%%%%%%%%%%%%%%%%%%%%%%%%%%%%%%%%%%%%%%%%%%%%%%%%%%%%%%%%%%%%%%%%
The direct detection cross section for the dark matter candidate contains a spin-independent component that is suppressed by the square of the perpendicular component of the velocity of the recoiling nucleus (\ie $v_\perp^2 \equiv v^2 - q^2/4\mu^2$, where $q$ is the momentum transfer and $\mu$ the dark matter-nuclei reduced mass) and a spin-dependent component suppressed by the momentum transfer squared (see \eg\cite{Gelmini:2018ogy} for dark matter-nuclei differential cross section); thus one naturally expects direct detection constraints to be significantly reduced relative to the constraints from colliders. To verify this, we have computed the constraints from XENON1T and PandaX-II explicitly~\cite{Aprile:2017iyp,Cui:2017nnn}\footnote{As this manuscript was being finalized, the XENON collaboration released an updated bound which increases current constraints on the cross section by a factor of $\sim 7$ for small dark matter masses and $\sim 2$ for large dark matter masses~\cite{Aprile:2018dbl}.}. 

The cross section for this interaction has been calculated using the tools provided in~\cite{Anand:2013yka}. The bounds have been computed for each experiment by taking the efficiency function and exposure from~\cite{Aprile:2017iyp} and~\cite{Cui:2017nnn}, and by applying Poisson statistics. This analysis reproduces to an accurate degree the published limits on the spin-independent interaction. 

\begin{figure}[t]
\centering
\begin{tabular}{cc}
 \includegraphics[width=0.48\textwidth]{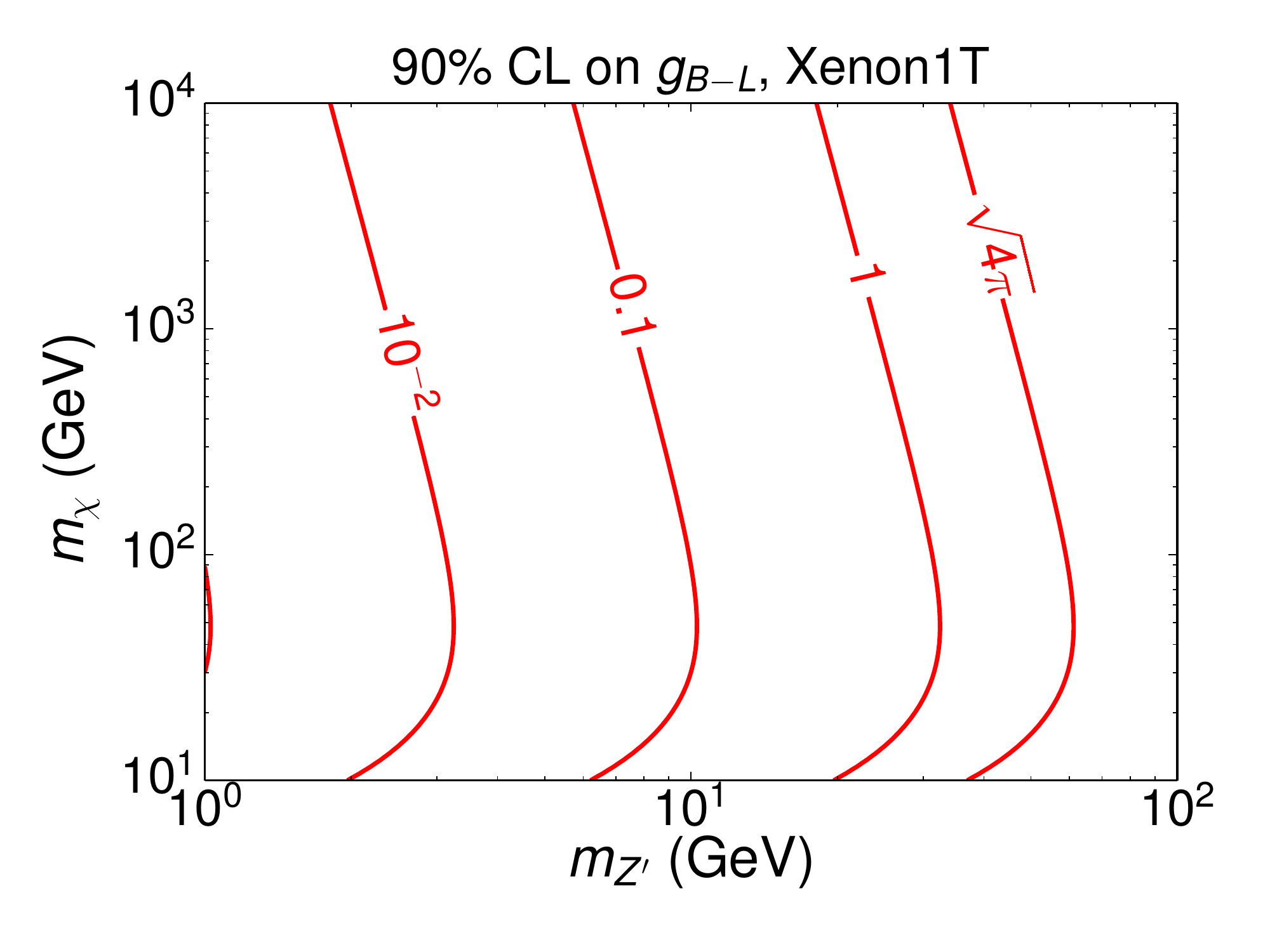} &  \includegraphics[width=0.48\textwidth]{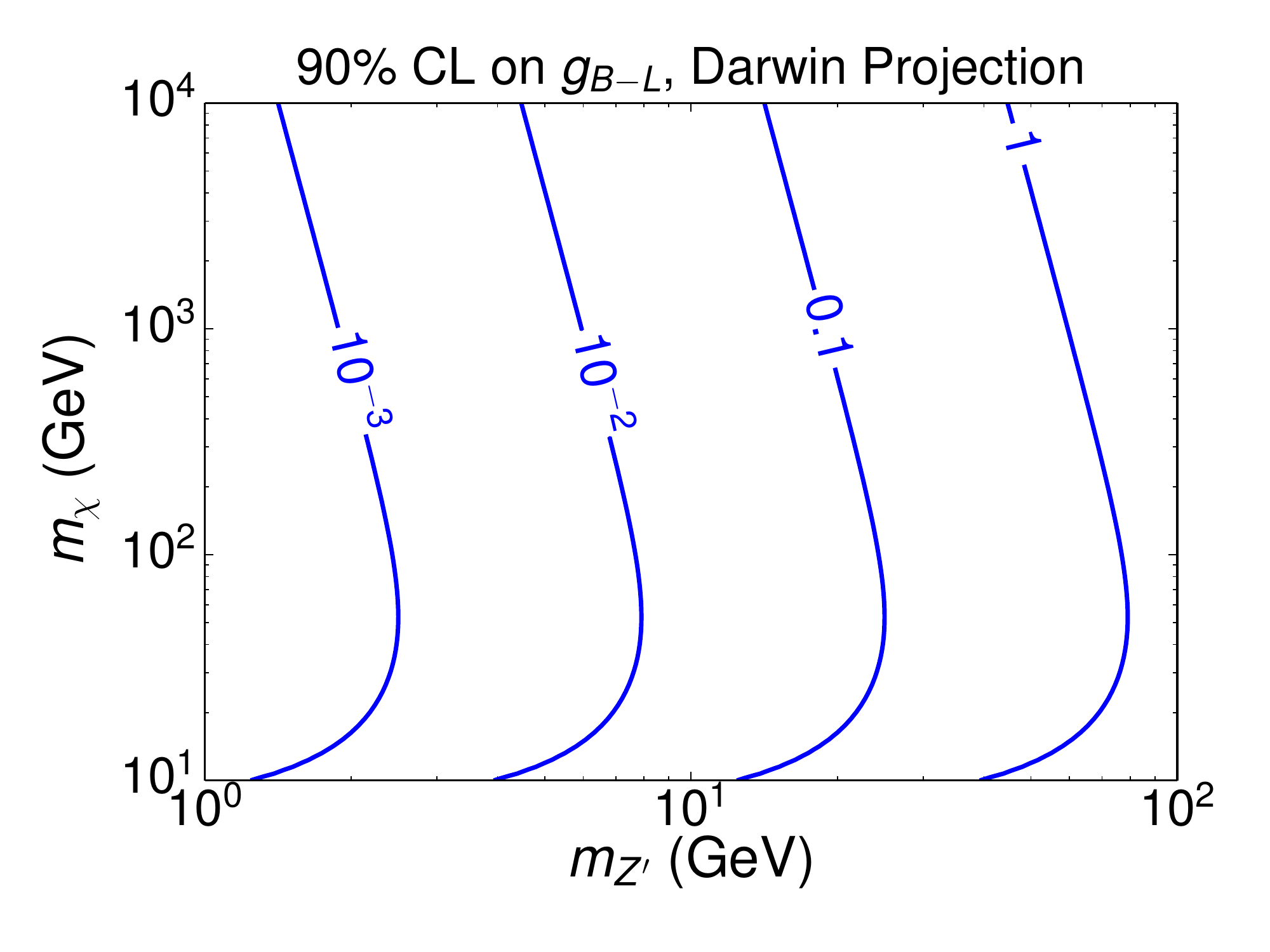}
 \end{tabular}
\caption{\textit{Left:} Direct detection constraints from Xenon1T at the $90\%$ CL. \textit{Right:} Projected sensitivity for Darwin assuming a 200 ton-year exposure and no observed events. }\label{fig:DD}
\end{figure}

The constraints on the coupling as a function of dark matter and $Z^\prime$ mass from the Xenon1T experiment are shown in \Fig{fig:DD}. We note that PandaX-II produces nearly identical limits. As expected, they are reduced with respect to collider searches in the parameter space of interest. 

In order to illustrate the sensitivity of future direct detection experiments, we also derived a 90\% CL sensitivity limit for the DARWIN experiment~\cite{Schumann:2015cpa}, assuming no events are observed (see right panel of \Fig{fig:DD}). This sensitivity limit is constructed considering perfect detection efficiency for nuclear recoils between 5 keV and 40 keV, a 40 ton fiducial volume and a 5 year runtime.

In our analysis we take the scalar-Higgs mixing to be zero; note that this is valid from a phenomenological perspective as this mixing is significantly constrained from direct detection searches and LHC Higgs measurements~\cite{Belanger:2013xza,Brehmer:2015rna,Beniwal:2015sdl,deBlas:2016ojx,Khachatryan:2016vau}. We have included for completeness the direct detection cross section arising from Higgs-mixing in the Appendix~\ref{subsec:Xsecs}.

%%%%%%%%%%%%%%%%%%%%%%%%%%%%%%%%%%%%%%%%%%%%%%%%%%%%%%%%%%%%%%%%%%%%%%%%%%%
\subsection{Full Parameter Space Scan}\label{subsec:DMscan}
%%%%%%%%%%%%%%%%%%%%%%%%%%%%%%%%%%%%%%%%%%%%%%%%%%%%%%%%%%%%%%%%%%%%%%%%%%%

Here, we describe the sampling method that allows for a comprehensive understanding of the viable parameter space without relying on simplifying assumptions that may bias the results. 

This parameter space scan begins by discretely sampling the $Z^\prime$ mass at 200 logarithmically spaced points between 1 GeV and 30 TeV. Note that we have truncated our analysis at $1$ GeV as strong constraints exist for MeV scale $Z^\prime$s that couple to both baryons and leptons (see \eg Ref.\cite{Ilten:2018crw})\footnote{Additionally, constraints from BBN require $m_{Z^\prime} \gtrsim \mathcal{O}(5)$ MeV~\cite{Boehm:2013jpa}.}.  For each fixed value of $m_{Z^\prime}$, randomly generated points are sampled in the 3-dimensional parameter space defined by ($m_\chi, m_{N_1}=m_{N_2}, m_\rho$), with all masses restricted to reside between $0.1$ GeV and $30$ TeV.  The $B-L$ gauge coupling is chosen so as to produce the correct relic abundance. A summary of the range of sampled parameters is contained in \Tab{tab:Parameters}. Since the interaction rate scales $\propto g_{B-L}^4$, this scan strategy is guaranteed to find the unique coupling that gives rise to the correct relic abundance. The chosen values of the parameters are only retained should they pass the following tests: \emph{(i)} the gauge coupling needed to produce the correct relic density is allowed by current constraints (see \Fig{fig:Zonly}) and \emph{(ii)} the values of the couplings remain perturbative (the perturbative limit is taken here to be $\sqrt{8\pi}$ for the Yukawas and $\sqrt{4\pi}$ for the $B-L$ gauge coupling). This sampling procedure is run continuously for $24$ hours, resulting in a total of $\sim7.5 \times 10^5$ points that are capable of providing a viable dark matter candidate. We emphasize that for a large region of parameter space, namely 150 GeV $\lesssim m_{Z^\prime} \lesssim 3$ TeV, there did not exist a single point passing the scan; the strength of the LHC $pp\rightarrow \ell^+\ell^-$ bound accounting for this exclusion has recently been pointed out in~\cite{Okada:2018ktp}.

\begin{table}[t]
\begin{center}
\begin{tabular}{ccc}
\hline\hline
  Parameter       &		 Description			 & 		Range     	       \\
\hline
  $m_\chi$         & Dark matter mass 	 & 	  $[0.1-3\times10^4]$ GeV    \\
    $m_{N_1}=m_{N_2}$         & Sterile Neutrino mass  			 & 	  $[0.1-3\times10^4]$ GeV     \\
    $m_{\rho}$&  Scalar mass &   $[0.1-3\times10^4]$ GeV   	  \\
    $m_{Z'}$&  $Z'$ mass &   $[1-3\times10^4]$ GeV   	  \\
    $g_{B-L}$& $B-L$ coupling &   $10^{-5}-\sqrt{4\pi}$   	  \\
  \hline \hline
\end{tabular}
\end{center}
\caption{Ranges of the parameters explored in the parameter scan. As discussed in \Sec{subsec:DMscan}, values of $m_{Z^\prime}$ were sampled using $200$ logarithmically spaced points. For each point random values of $m_\chi, \, m_{N_1}=m_{N_2}, \, m_{\rho}$ were drawn from the range detailed in the table above (assuming a log-flat distribution), and $g_{B-L}$ was set to be consistent with the measured relic abundance at $2\sigma$, \ie $\Omega_{DM} h^2 = 0.120 \pm 0.003$~\cite{Ade:2015xua}. We further require all Yukawa couplings to be $\leq \sqrt{8\pi}$ and $g_{B-L} \leq \sqrt{4\pi}$ in order for the model to remain perturbative. The total number of points in parameter space that meet such requirements are $7.5 \times 10^5$. }\label{tab:Parameters}
\end{table}
One might wonder whether a slight reduction of the constraints arising from a partial branching fraction to invisible final states might significantly alter the results of this initial parameter scan. This would be the case \eg in models with a Majorana dark matter containing lepton number $L>1$.
 In order to investigate this possibility we repeat the scan, but only retaining values of $g_{B-L}$ that reside between the current bound $g_{B-L}^{\rm lim}$ and the reduced bound, where the reduced bound is 
  defined as $g_{B-L}^{\rm lim} / 0.7$\footnote{This factor has been conservatively chosen by analyzing the modification to the DY dilepton spectra in {\tt MadGraph5 aMC@NLO}, in order to reflect the strength of the bound that would be derived should the maximally allowed invisible branching ratio be realized.}. In this case, an additional constraint is imposed that requires either $m_\chi$ or $m_{N_1,N_2}$ $< m_{Z^\prime}/2$, since these are the decays responsible for the reduced bounds.

The most important results of the parameter space scan can be summarized in the following:
\begin{itemize}
\item if the dark matter mass $m_\chi \lesssim 900$ GeV, the model only remains viable if annihilations are on-resonance;
\item strong constraints from high energy dilepton searches rule out the model for $150$ GeV $\lesssim m_{Z^\prime} \lesssim 3$ TeV, even for annihilations proceeding through a resonance;
\item as will be discussed shortly in \Sec{sec:ThC}, the viable parameter space passing resonance cuts for $m_\chi \lesssim 5$ TeV and $m_{Z^\prime} \lesssim 4$ TeV often have deep theoretical flaws arising from the appearance of low-energy Landau poles or an instability of the potential.
\end{itemize}
We defer a more comprehensive discussion of the parameter scan results to \Sec{sec:ThC}.

%%%%%%%%%%%%%%%%%%%%%%%%%%%%%%%%%%%%%%%%%%%%%%%%%%%%%%%%%%%%%%%%%%%%%%%%%%%
\subsection{Annihilation Phenomenology}\label{subsec:annhil}
\begin{figure}[t]
\centering
\begin{tabular}{ccc}
   p-wave &  s-wave &   s-wave   \\
   \includegraphics[width=0.30\textwidth]{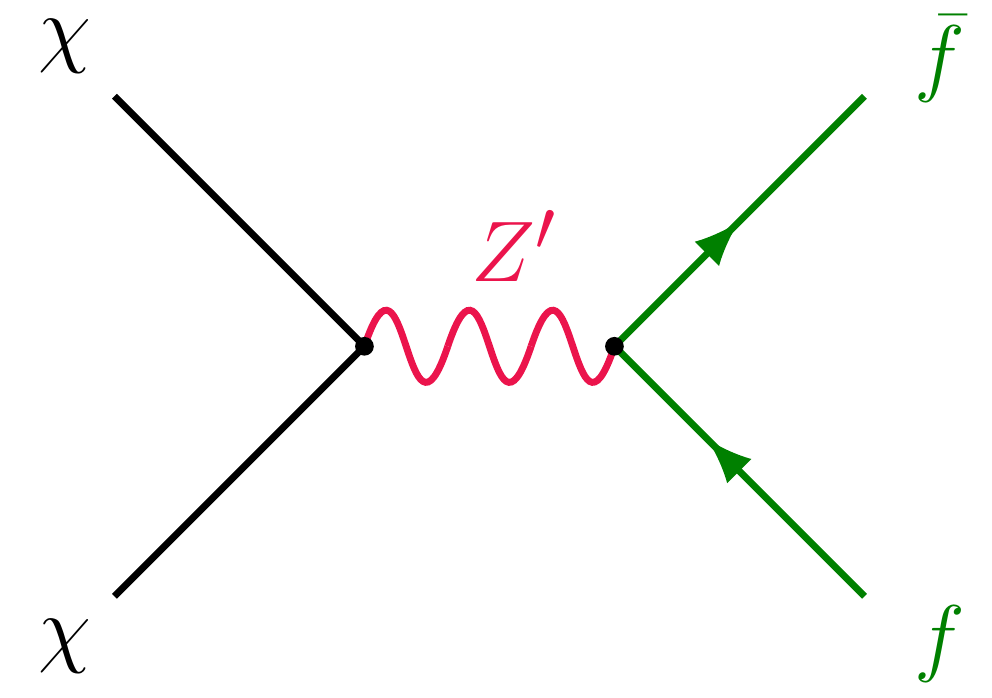} &  \includegraphics[width=0.30\textwidth]{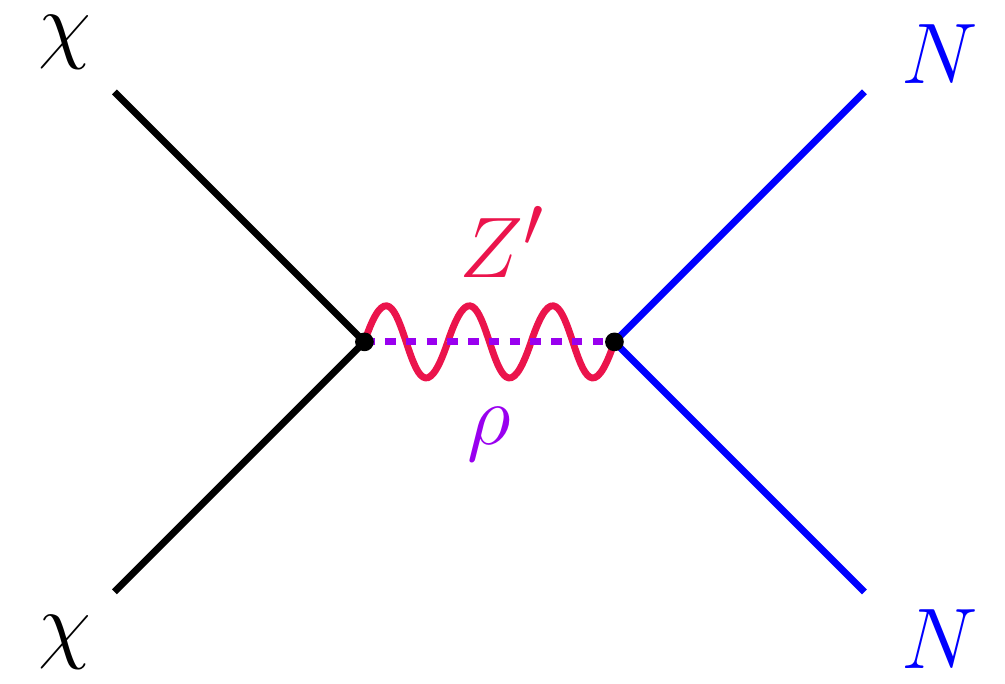} &   \includegraphics[width=0.30\textwidth]{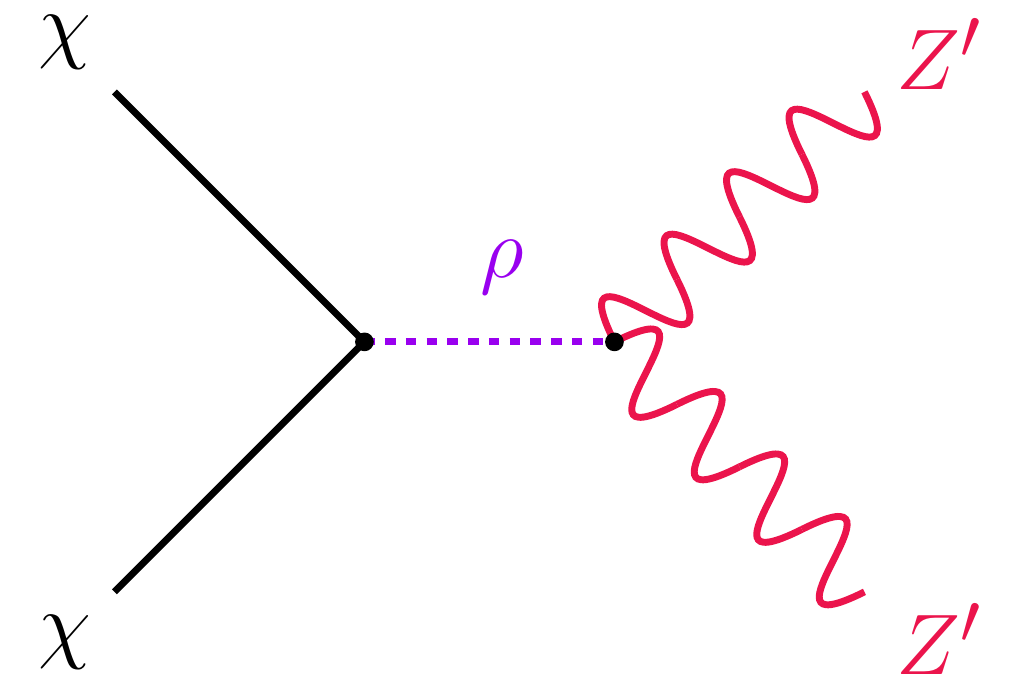}   \\
   \end{tabular}
\begin{tabular}{ccc}
   p-wave &  s-wave &   s-wave   \\
   \includegraphics[width=0.30\textwidth]{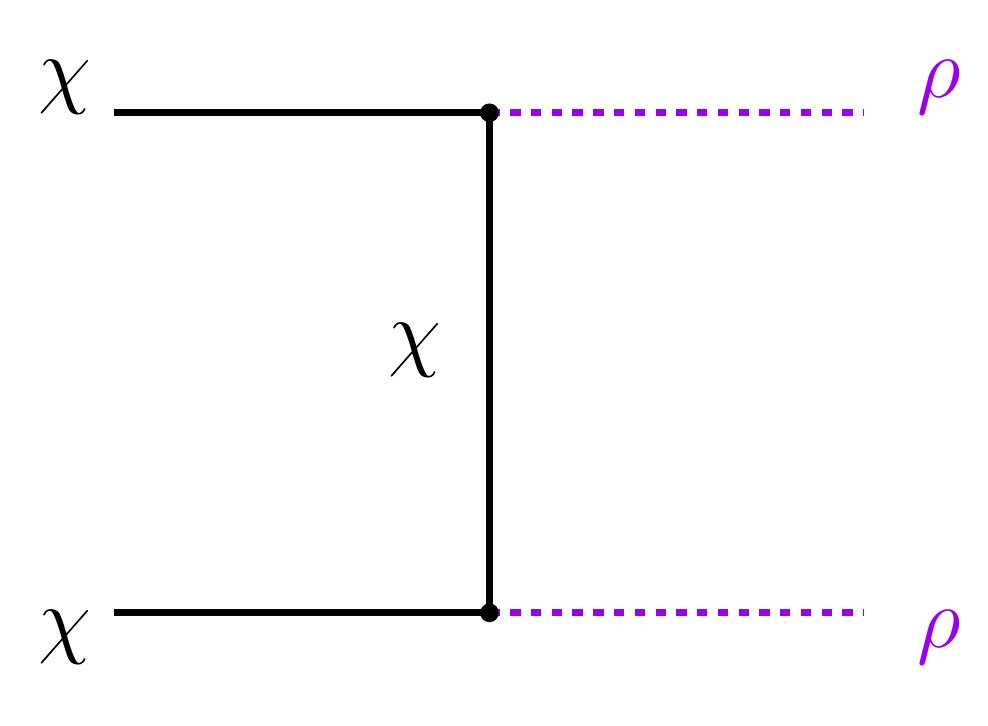} &  \includegraphics[width=0.30\textwidth]{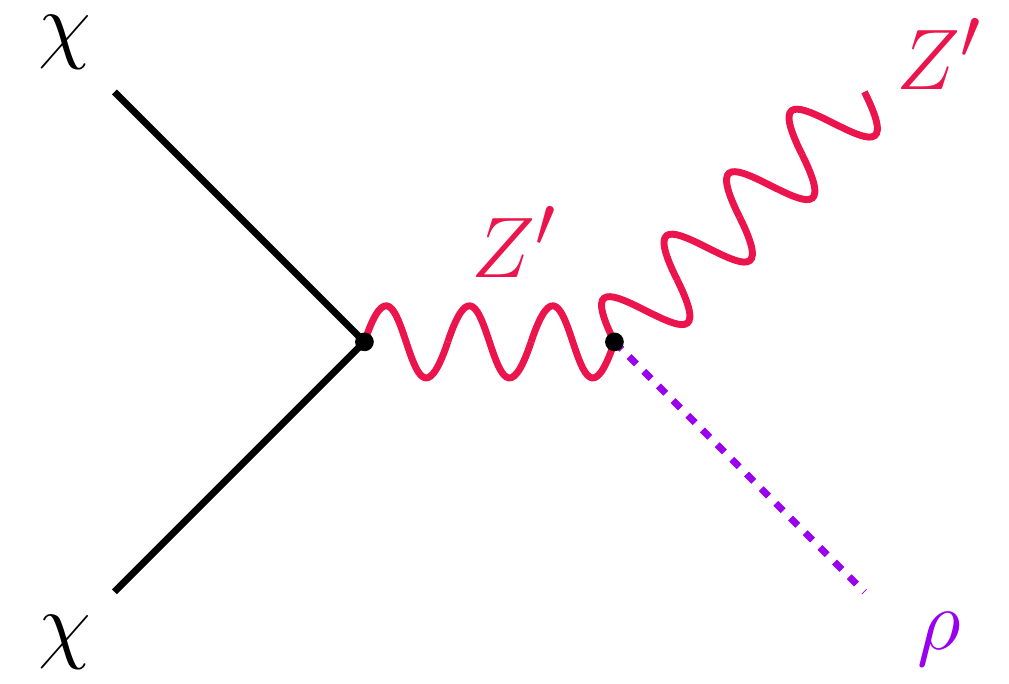}   &   \includegraphics[width=0.30\textwidth]{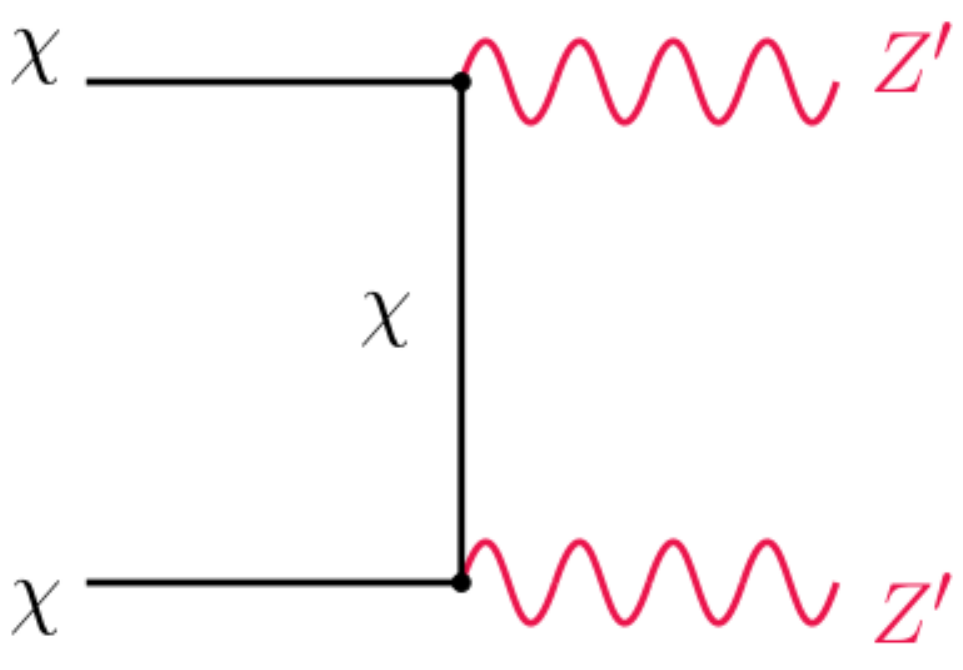}  \\
   \end{tabular}
\caption{Representative sample of relevant dark matter annihilation diagrams.}\label{fig:diagrams}
\end{figure}

%%%%%%%%%%%%%%%%%%%%%%%%%%%%%%%%%%%%%%%%%%%%%%%%%%%%%%%%%%%%%%%%%%%%%%%%%%%
By analyzing the viable parameter space identified with the scan described in \Sec{subsec:DMscan}, we can determine the relevant annihilation phenomenology in various regimes. We begin by presenting some of the relevant annihilation channels in \Fig{fig:diagrams} -- this is not intended to be a comprehensive list but rather it serves as a representative sample. In \Fig{fig:diagrams}, we have also listed the annihilation cross section's leading order behavior in velocity. Assuming the dark matter abundance does not proceed through a resonance, the annihilation phenomenology can generically be summarized with two points: \emph{(i)} if kinematically accessible, then the annihilation proceeds to $Z'\rho$, and \emph{(ii)} if this channel is not accessible, the annihilation predominantly to $NN$. Both channels are s-wave and lead to SM final states that are detectable by indirect searches.
 
In order to highlight some of the most important features of the annihilation phenomenology for the viable parameter space, we identify three distinct regimes that are differentiated by the mass of the $Z'$ and motivated by the weakness of collider constraints in those regimes; namely, we consider $1 \, \text{GeV} <m_{Z'} < 10 \, \text{GeV}$, $m_{Z'} \sim m_{Z}$ and $m_{Z'} > 3 \, \text{TeV}$.

\begin{enumerate}
\item $1 \, \text{GeV} <m_{Z'} < 10 \, \text{GeV}$. Due to the strong background that the $\phi(1020), \, J/\psi$ and $ \psi(2S)$ resonances produce in the BaBar search~\cite{Lees:2014xha} their surrounding ($\pm 30-50 \, \text{MeV}$) regions were eliminated from their study. However, their 2009 search did not perform this cut~\cite{Aubert:2009cp}, and although weaker, there do exist constraints in such region (as shown in Figure~\ref{fig:Zonly}). For each case we detail what are the allowed couplings, their corresponding dark matter masses, and the dominant annihilation channel:

\begin{enumerate}
\item $m_{Z'} \sim m_{\phi(1020)}$. We have found that for dark matter masses greater than 50 GeV, the correct relic abundance is obtained with $g_{B-L} \simeq 2\times10^{-3}\left(\frac{50 \, {\rm GeV}}{m_\chi}\right)$ and annihilations in the contemporary Universe proceed to either $Z' \rho$ or $NN$ with an annihilation rate of $\left<\sigma v \right> \sim 2 \times 10^{-26} \text{cm}^3/\text{s}$. Thus, limits from dark matter annihilation in dwarf galaxies~\cite{Fermi-LAT:2016uux} rule out $m_{\chi} \lesssim 100 \, \text{GeV}$, or equivalently $g_{B-L} \gtrsim 9\times 10^{-4}$.
\item $m_{Z'} \sim m_{J/\psi}, \,  m_{\psi(2S)},\, m_{\Upsilon}$. Using the arguments outlined above for $m_{Z'} \sim m_{\phi(1020)}$, one can state that obtaining the correct relic abundance requires $m_\chi/\text{TeV} > 1 , \, 12, \, 0.6$, for the $m_{J/\psi}, \,  m_{\psi(2S)},$ and $\, m_{\Upsilon}$ resonances, respectively. 

\end{enumerate}

\item $m_{Z'} \sim m_{Z}$. Constraints derived in this region in parameter space are strongly suppressed due to the large backgrounds from the SM $Z$ (as shown in Section~\ref{subsec:LHC}). However, this region of parameter space can potentially be tested with indirect searches. For $ 86 \, \text{GeV} \lesssim m_{Z'} \lesssim 97 \, \text{GeV}$ and $m_\chi \gtrsim 500 \, \text{GeV} $, the annihilation cross section is $\left<\sigma v \right> \sim 2 \times 10^{-26} \text{cm}^3/\text{s}$, and proceeds either to $Z' \rho$ (if $2 m_{\chi} > m_{Z'}+m_\rho$), or to $NN$ if the former is kinematically forbidden. Although this region of parameter space is currently unconstrained by indirect searches, future searches could potentially probe this region. It is also worth mentioning that a dedicated study of the the dilepton search channel probing invariant mass distributions close to the mass of the SM $Z$ could conceivably test this region, since relatively large gauge couplings (\ie $ g_{B-L} \simeq 0.034 \frac{500 \, \text{GeV}}{m_{\chi}}$) are required to produce the relic abundance.

\item $m_{Z'} > 3 \, \text{TeV}$.
\begin{enumerate}
\item $2 m_\chi > m_\rho + m_{Z'}$. Since the $Z' \rho$ channel is accessible, it will always dominate, and if $2 m_\chi > 1.5\left(m_\rho + m_{Z'}\right)$ the annihilation rate relevant for indirect searches is $\left< \sigma v \right> > 10^{-26} \text{cm}^3/\text{s}$, and thus potentially testable with indirect observations from future ground-based observatories.

\item $2 m_\chi < m_\rho + m_{Z'}$. If the $Z'\rho$ channel is closed, the annihilation final state can be $\rho \rho$, $Z'Z'$, $\bar{f}f$ and $NN$, depending on the masses of these particles. 
In particular if $g_{B-L} < 1$, then the annihilation proceeds predominantly to $NN$ via $\rho$ exchange, or $\bar{f} f$ via resonant $Z'$ exchange; the $Z'Z'$ channel, on the other hand, only occurs in the resonant region $m_\chi \sim m_{\rho}/2$. We note that the annihilation to SM fermions is p-wave suppressed, while the annihilation to $NN$ is s-wave, but when mediated by the $\rho$ is Yukawa suppressed. Thus, only if $m_\rho \gtrsim m_\chi$ and $m_N \lesssim m_\chi$ the annihilation rate will be comparable to $\left<\sigma v \right> \sim 10^{-26} \text{cm}^3/\text{s}$, which represents a very small portion of parameter space. This can explicitly be seen in Figure~\ref{fig:TeV_param}.
\end{enumerate}

\end{enumerate}

%%%%%%%%%%%%%%%%%%%%%%%%%%%%%%%%%%%%%%%%%%%%%%%%%%%%%%%%%%%%%%%%%%%%%%%%%%%
\section{Summary \& Theoretical Considerations}\label{sec:ThC}
%%%%%%%%%%%%%%%%%%%%%%%%%%%%%%%%%%%%%%%%%%%%%%%%%%%%%%%%%%%%%%%%%%%%%%%%%%%

Thus far we have identified the parameter space for which collider constraints on the $U(1)_{B-L}$ gauge boson do not exclude the possibility that one of the right-handed neutrinos constitutes the dark matter of our Universe. The important question that remains is whether or not the viable parameter space is theoretically well-motivated. In this section we attempt to address exactly this question. Here, we show that much of viable parameter space exists only on-resonance, and most of the viable parameter space not on-resonance exhibits deeper theoretical issues related to the appearance of a Landau pole or an instability of the scalar sector at remarkably low energies, which become apparent only through an analysis of the RGEs.

We begin by considering how many of the viable points identified in the parameter scan require production through a resonance. Models annihilating through the $Z^\prime$ or $\rho$ resonance present the most likely candidates for surviving this analysis, and thus we begin here. For $m_\chi \leq 900$ GeV, the requirement that $m_\chi \geq 0.6 m_{Z^\prime}$ or $m_\chi \leq 0.4 m_{Z^\prime}$ removes more than 70\% of the originally identified viable points. Further imposing that $m_\chi \geq 0.6 m_{\rho}$ or $m_\chi \leq 0.4 m_{\rho}$ removes an additional $\sim23\% $. A quick look at the remaining parameter space reveals that the mass of the $Z^\prime$ must be near the mass of the $Z$ (\ie $m_{Z^\prime}\sim m_Z$), a consequence of the fact that collider constraints are weakened near the mass of the $Z$. Imposing the requirement that $m_{Z^\prime} > 100$ GeV or $m_{Z^\prime} < 80$ GeV removes all remaining points. Thus, one can concretely say that for dark matter masses $m_\chi \leq 900$ GeV the only viable parameter space is produced via a resonance, or requires the $Z^\prime$ to be nearly degenerate with the SM $Z$.

One can perform a similar analysis as above, systemically removing resonance points, but for larger values of the dark matter mass. We have performed this analysis for two independent regions, $900$ GeV $\leq m_\chi \leq 5$ TeV and $5$ TeV $\leq m_\chi \leq 6$ TeV; removing these resonances eliminates $\sim 55\%$ and $\sim 20\%$ of the originally identified viable points, for each region respectively (a summary of these cuts is provided in the top half of \Tab{table:adaf}).

The more conceptional issues we address in this section deal with the appearance of a Landau pole in the running of the $B-L$ gauge coupling, and the stability of the scalar sector, both of which require analyzing the RGEs of this model.

The two-loop RGEs are calculated using {\tt SARAH}~\cite{Staub:2008uz,Staub:2015kfa}. For simplicity, only the one-loop RGEs for the gauge coupling, the quartic coupling of the new scalar, and the couplings of the right-handed neutrinos to the new scalar are provided below\footnote{Note that in principle there exist a total of three gauge couplings that enter a model with two $U(1)$ symmetries, the extra coupling arising from the potential kinetic mixing term. We have implicitly assumed in the above discussion that this mixing is effectively zero, as non-zero $Z-Z^\prime$ constraints can be stringent (see \eg\cite{Hook:2010tw,Chun:2010ve,Mambrini:2011dw,An:2014twa,Abdullah:2018ykz}). However, in principle, the RGEs may evolve so as to induce a kinetic mixing at higher energies; \eg defining the mixing to vanish at a scale $\mu(\epsilon = 0)$ induces a non-zero mixing of magnitude $\epsilon(\mu) \sim 0.015\times g_{B-L}\log\frac{\mu(\epsilon=0)}{\mu}$. We have verified that for this model, and for the energies of interest, this does not happen. Thus for simplicity we have set this kinetic mixing term in the RGEs shown in \Eq{eq:rges} to zero, however its evolution has been included in the calculations.}:
\begin{subequations}\label{eq:rges}
\begin{align}
& \beta_{g_{B-L}}^{(1)}   =  
12 g_{B-L}^{3} m_{Z^\prime} \,,  \\
&\beta_{\lambda_\phi}^{(1)}  =  
2 \Big(48 g_{B-L}^{4}  + 10 \lambda_{\phi}^{2}  + 2 \lambda_2 \left(\lambda_\chi^2+\lambda_{N_1}^2+\lambda_{N_2}^2\right)  -2 \left(\lambda_\chi^4+\lambda_{N_1}^4+\lambda_{N_2}^4\right) -24 g_{B-L}^{2} \lambda_\phi  + \lambda_{H\phi}^{2}\Big)\,,   \\
& \beta_{\lambda_\chi}^{(1)}  =  
2 \lambda_\chi^3  + \lambda_\chi \Big(-6 g_{B-L}^{2} +  \left(\lambda_\chi^2+\lambda_{N_1}^2+\lambda_{N_2}^2\right)\Big)\,,    \\
& \beta_{\lambda_{N_{1,2}}}^{(1)}  =  
2 \lambda_{N_{1,2}}^3  + \lambda_{N_{1,2}} \Big(-6 g_{B-L}^{2} +  \left(\lambda_\chi^2+\lambda_{N_1}^2+\lambda_{N_2}^2\right)\Big) \, ,
\end{align}
\end{subequations}
where Yukawa contributions $Y_{\alpha\beta}$ have been set to zero as these are negligible. A more comprehensive discussion of the RGEs in the $U(1)_{B-L}$ model can be found in \eg~\cite{Basso:2010jm,Khoze:2014xha}. We define the couplings at the freeze-out scale, \ie $m_\chi/20$. If $m_{Z^\prime}$ is above this scale, the $B-L$ gauge coupling is not evolved below the mass of the $Z^\prime$. 
\begin{figure}[t]
\centering
\hspace{-0.1cm}\includegraphics[width=0.48\textwidth]{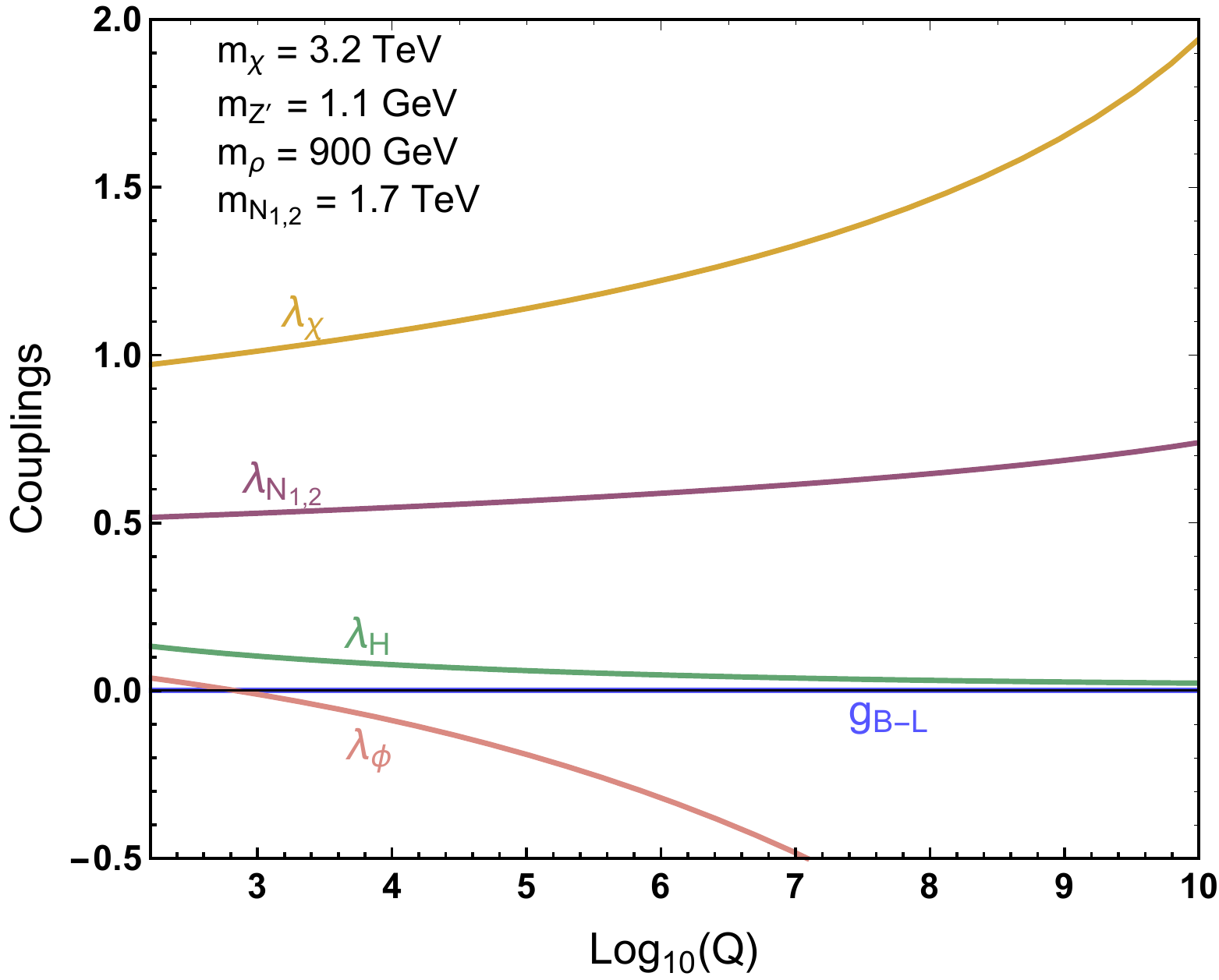}\hspace{0.4cm}
\includegraphics[width=0.48\textwidth]{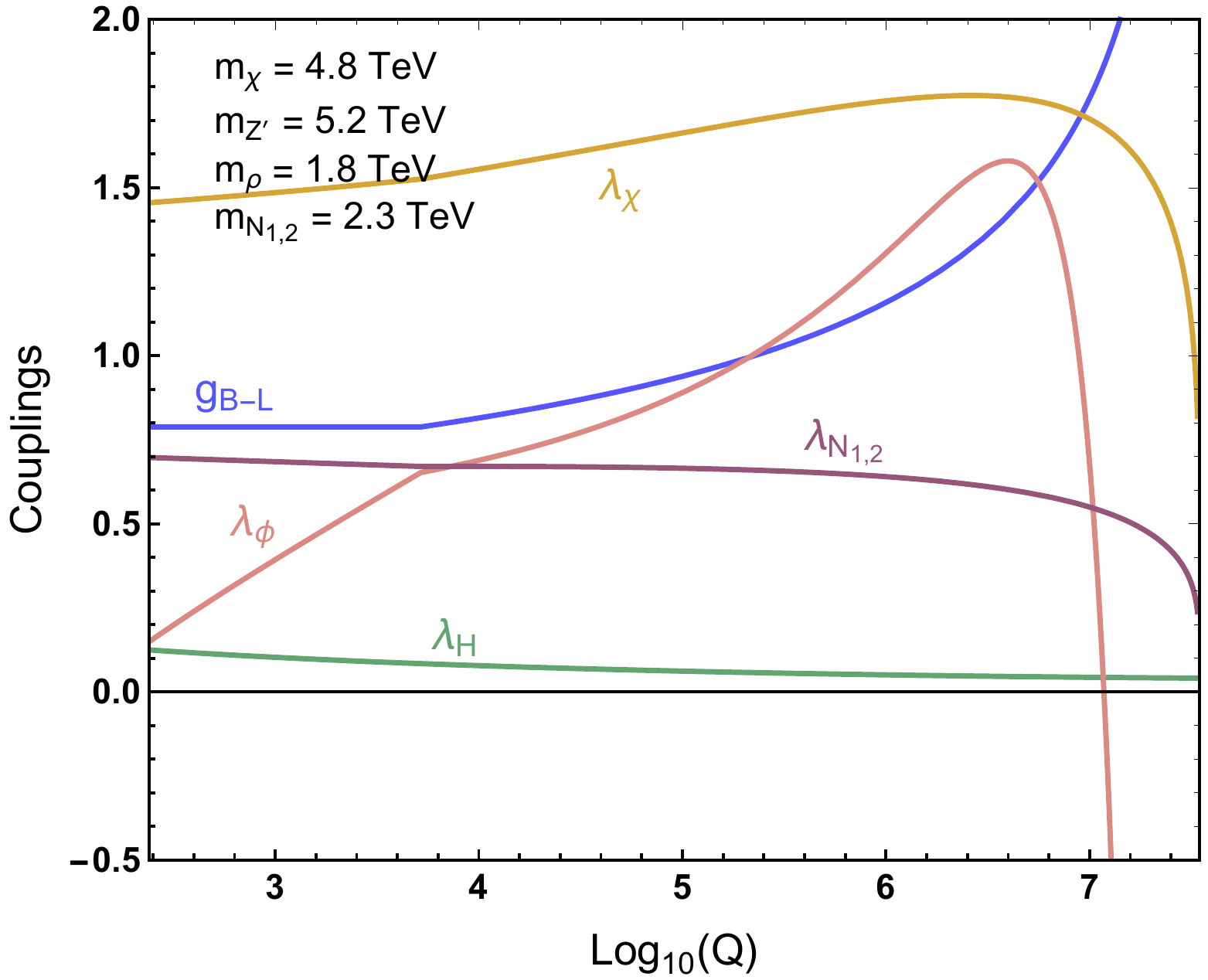}
\caption{\label{fig:egrge}Evolution of couplings for various choices of $m_\chi$, $m_{Z^\prime}$, $m_\rho$, and $m_{N_1,2}$. Masses are chosen to illustrate scalar instability (left) and the appearance of a Landau pole (right). The gauge coupling has been chosen so as to produce the correct relic density.}
\end{figure}

\begin{figure}
\centering
\includegraphics[width=0.495\textwidth,trim={0 0 0.2cm 0cm},clip]{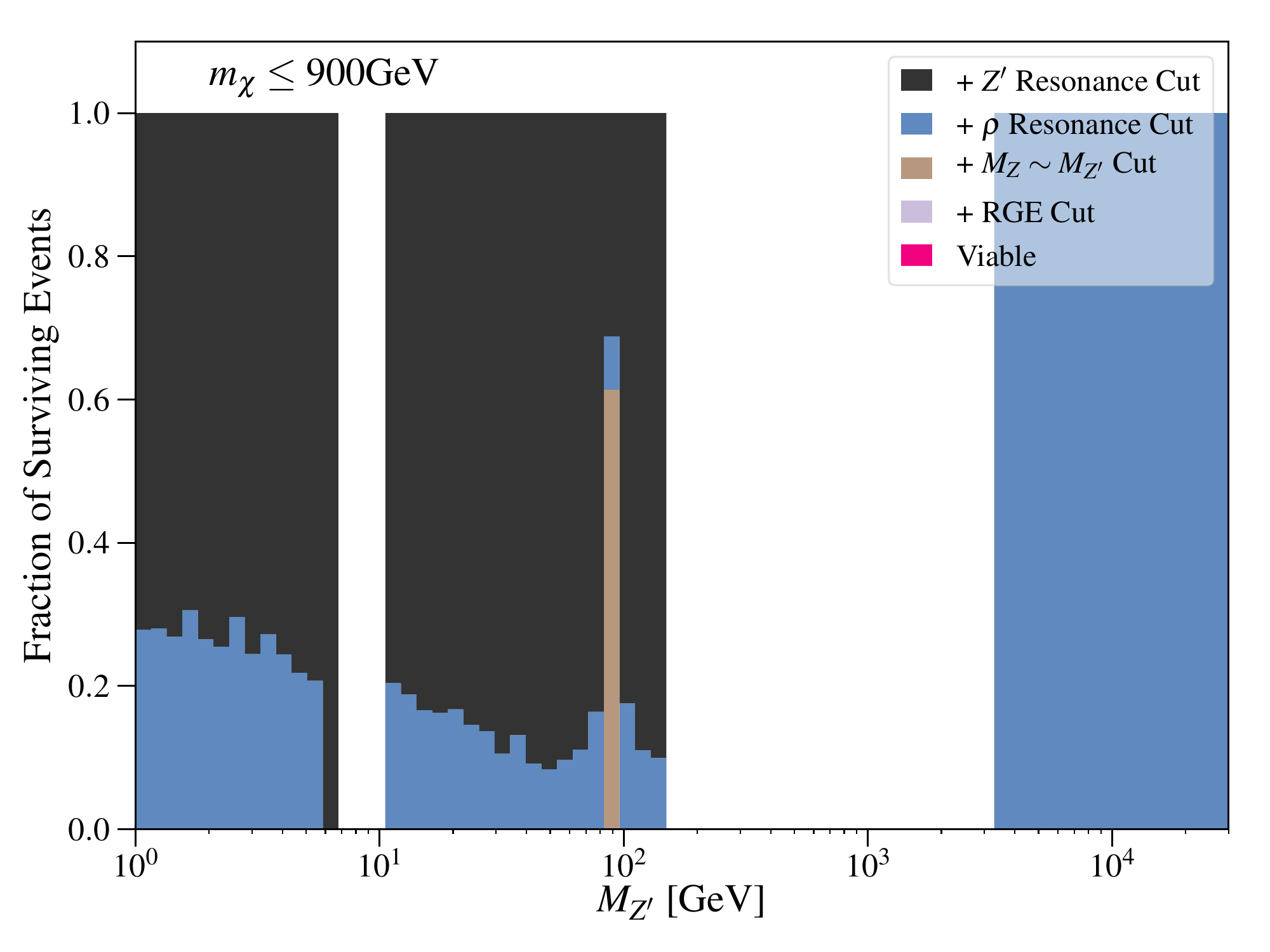}
\includegraphics[width= 0.495\textwidth,trim={0.2cm 0 0cm 0cm},clip]{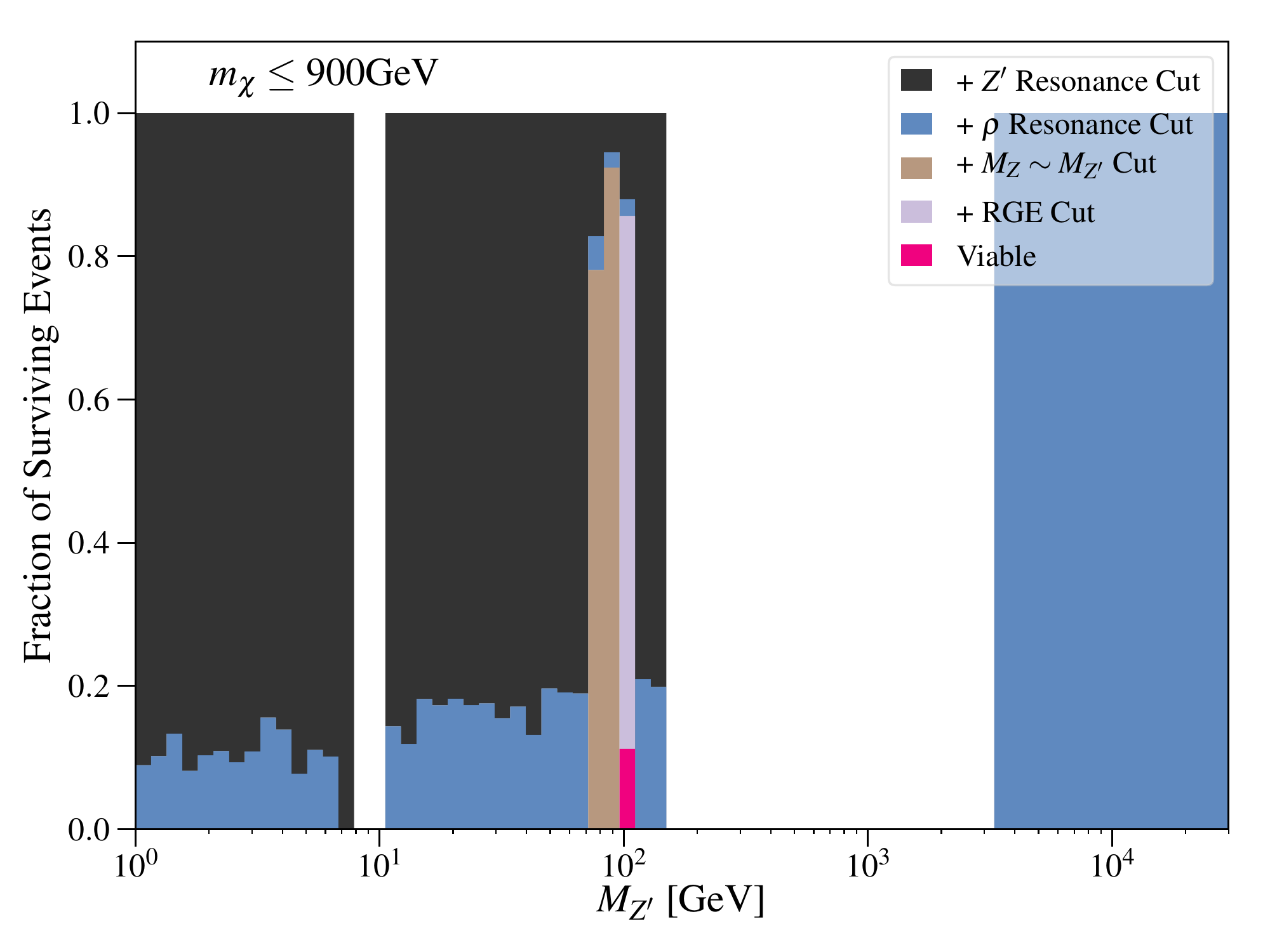}

\includegraphics[width=0.495\textwidth,trim={0 0cm 0.2cm 0},clip]{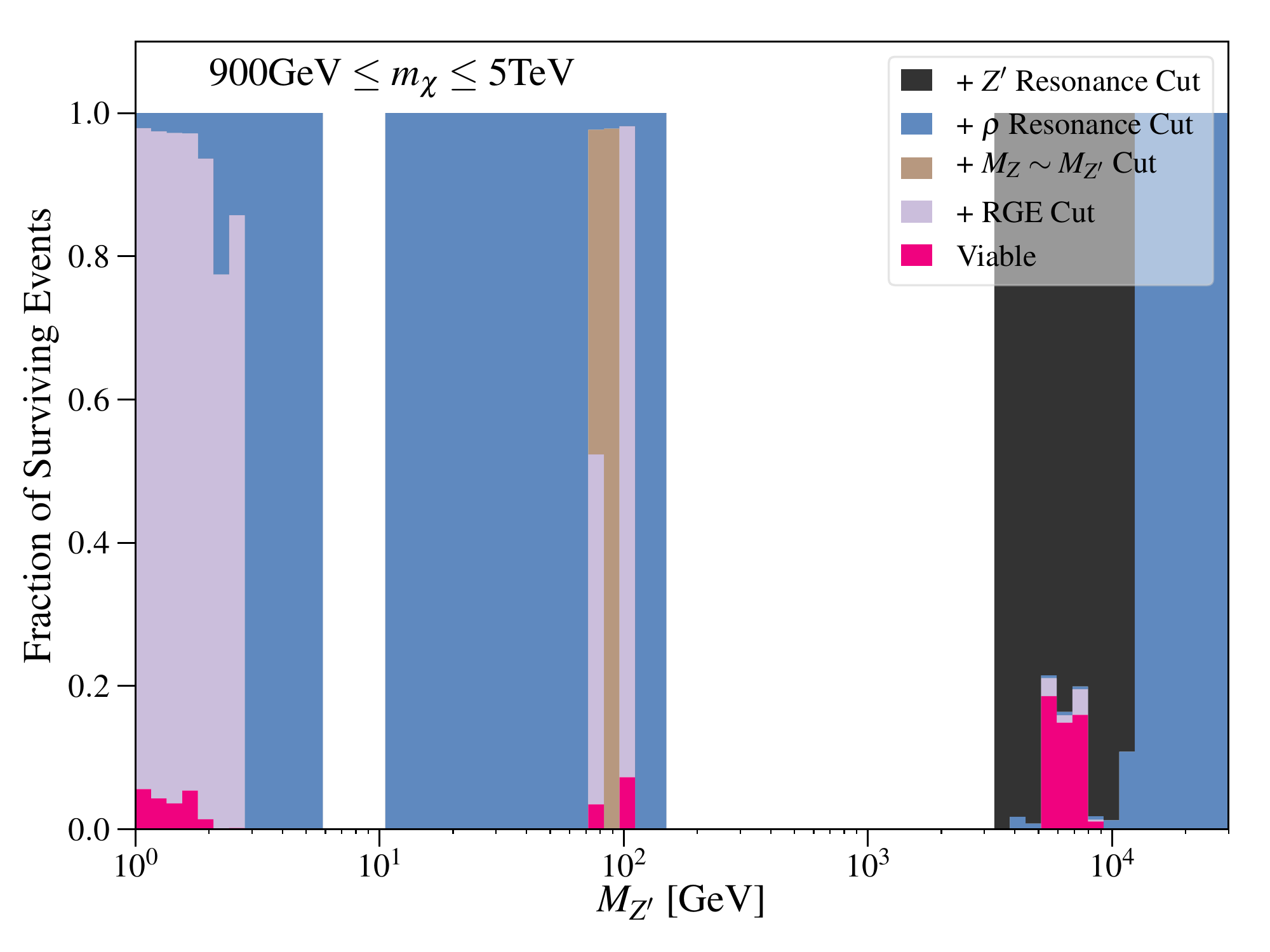}
\includegraphics[width= 0.495\textwidth,trim={0.2cm 0cm 0 0},clip]{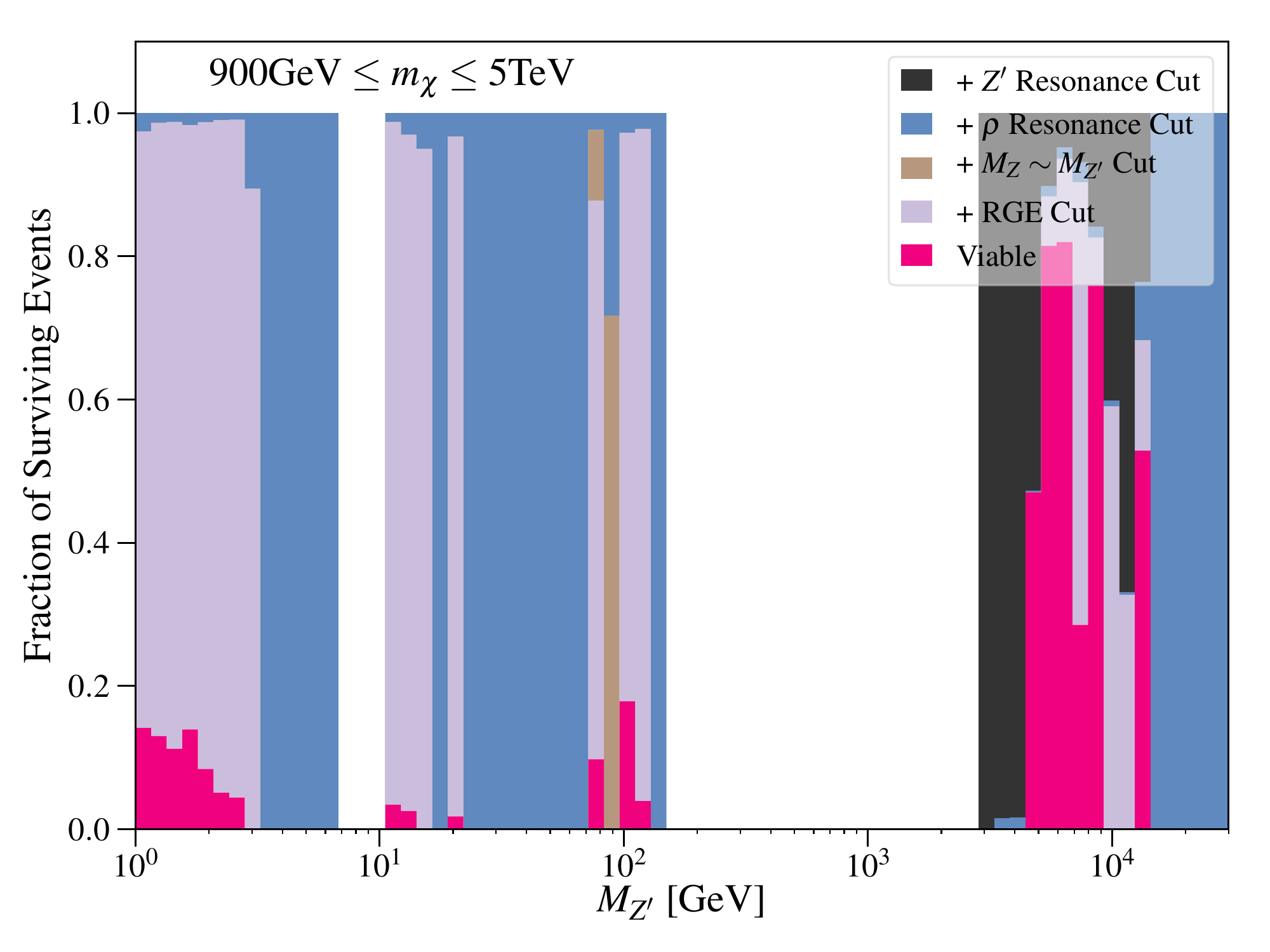}

\includegraphics[width=0.495\textwidth,trim={0cm 0cm 0.2cm 0},clip]{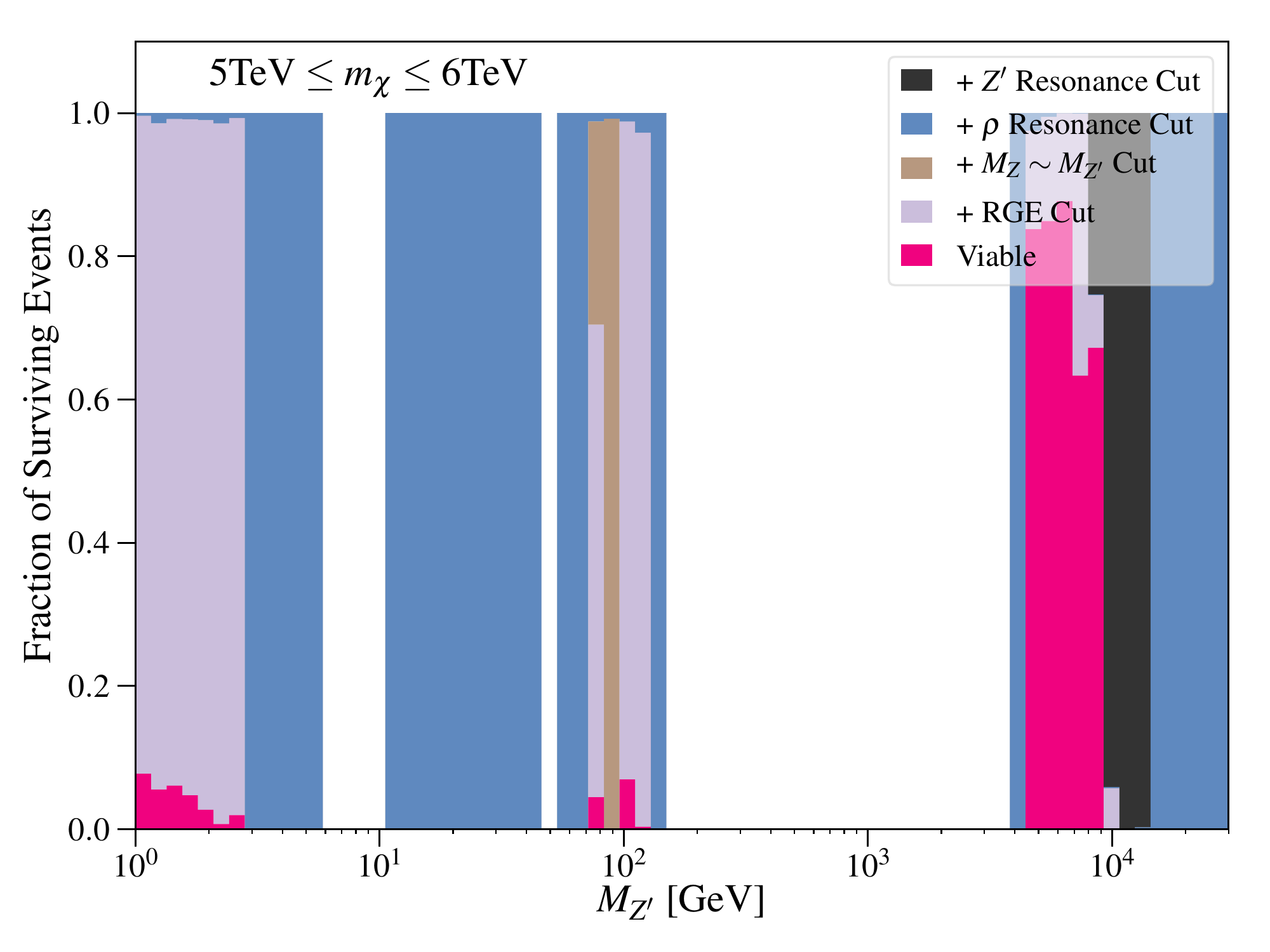}
\includegraphics[width= 0.495\textwidth,trim={0.2cm 0cm 0 0},clip]{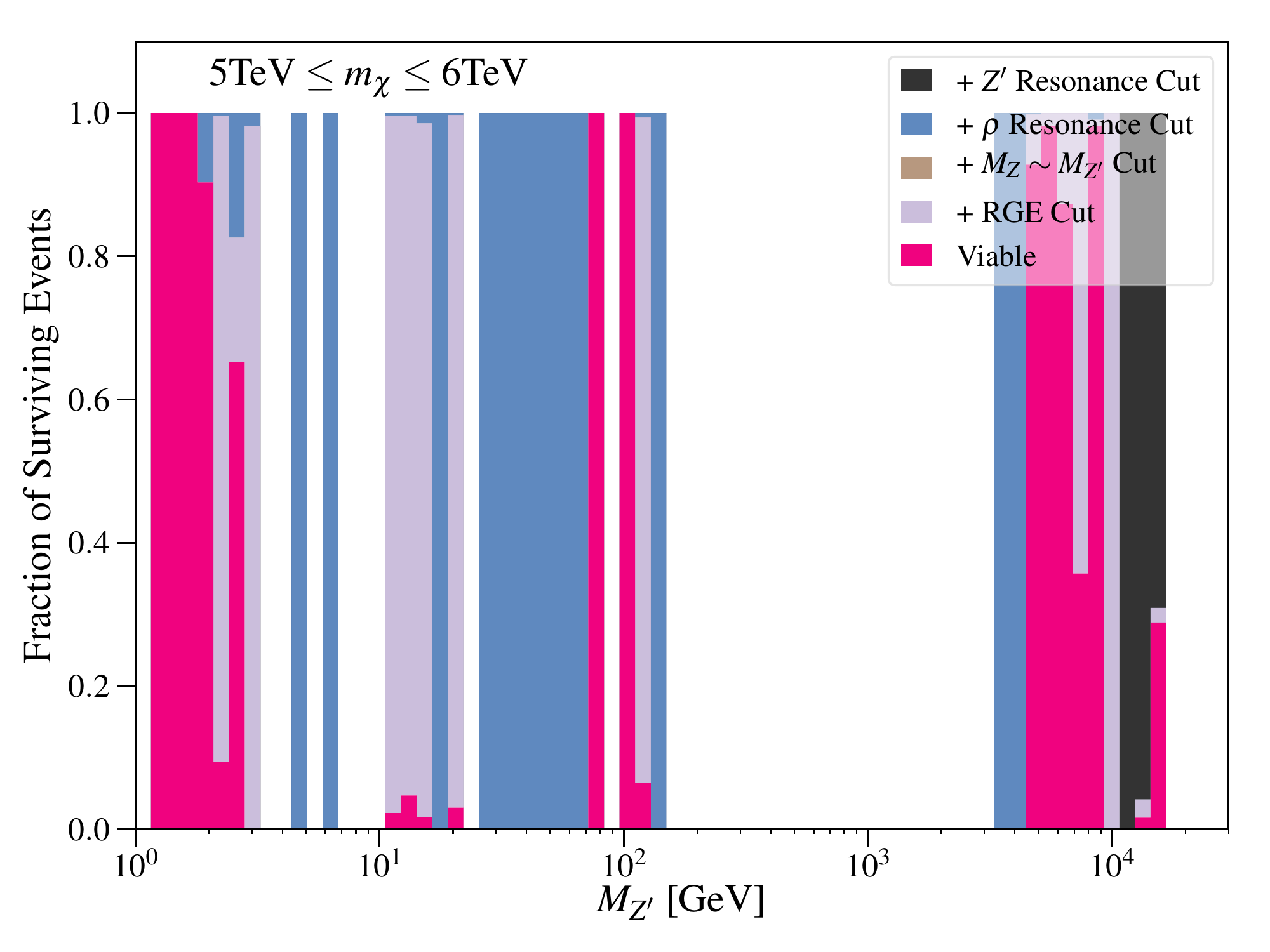}

\caption{Histograms showing the fraction of sampled points removed by each of the cuts described in \Sec{sec:ThC} for $m_\chi < 900$ GeV (top), 900 GeV $< m_\chi <$ 5 TeV (middle), and 5 TeV $< m_\chi<$ 6 TeV (bottom) as a function of $m_{Z^\prime}$. Left (right) panels denote the results for the full (reduced) bound analysis. Viable points surviving all cuts are shown in pink. When no histogram is shown (\eg $150\, {\rm GeV} < m_\chi < 3$ TeV) it is understood that no points in the sampling routine were capable of producing a viable dark matter candidate. Numerical values detailed the percentage of points surviving each cut are shown in \Tab{table:adaf}. \label{fig:histo1}}
\end{figure}

After preforming the aforementioned resonance cuts, we use the RGEs to determine at what energy either \emph{(i)} a Landau pole appears, \emph{(ii)} the scalar sector becomes unstable (\ie when either $4\lambda_{H}\lambda_\phi - \lambda_{H\phi}^2$ becomes negative or when $\lambda_H,\lambda_\phi < 0$), or \emph{(iii)} one of the couplings becomes non-perturbative (\ie exceeds $\sqrt{8\pi}$). If this energy is below ${\rm Max}(m_\chi, m_{Z^\prime}, m_\rho, m_{N})$, we deem this model to have an inconsistency and reject this point in parameter space. The results for these cuts, in the order they have been presented above, is listed in \Tab{table:adaf} and shown graphically in \Fig{fig:histo1}. In Fig.~\ref{fig:egrge} we show two examples of the RGEs that highlight the instability of the scalar sector (left) and the appearance of a Landau pole (right).
Notice that in models with a dark matter  lepton number $L \neq 1$, the running of the couplings will change, in particular $L > 1$ could aggravate the problem with the running of $g_{B-L}$, making such models inconsistent in a wider region of the parameter space.  

For dark matter masses in the range $900$ GeV $\leq m_\chi \leq$ 5 TeV, the perturbativity and scalar stability cut drastically reduce the viable parameter space for $m_{Z^\prime} \lesssim 5$ TeV; in particular, after all aforementioned cuts, only $\sim 7\%$ of the originally identified points remain. The same cut applied for the 5 TeV $\leq m_\chi \leq$ 6 TeV region similarly removes nearly all points, except for those in the range 5 TeV $\lesssim m_{Z^\prime} \lesssim$ 10 TeV (in total, roughly $30\%$ of the originally identified points remain). 

\begin{figure}[h]
\centering
\includegraphics[width=0.99\textwidth]{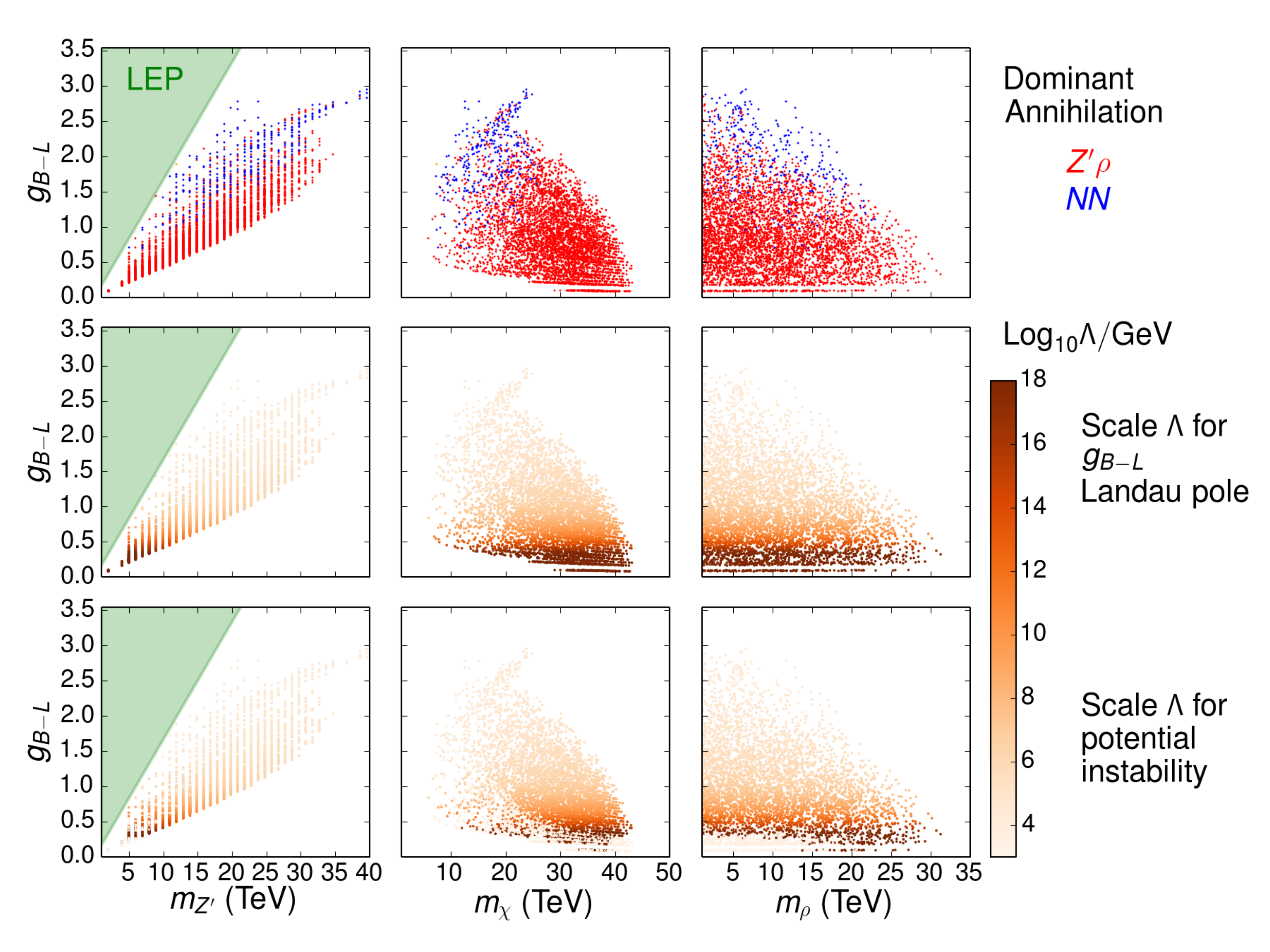}
\caption{High mass region of the model. All points correspond to $\Omega_\chi h^2 = 0.12$, resonant regions ($0.4<m_\chi/m_{Z'}, \, m_\chi/m_{\rho} < 0.6$) have been removed and perturbative unitarity on the couplings and masses has been imposed (see Eqs~\ref{eq:mass_pert}-\ref{eq:coupl_pert}). The \textit{upper panel} shows the parameter space categorized by the dominant annihilation channel in the contemporary Universe. The \textit{middle panel} shows the result of the parameter scan, with color coding depicting the scale at which a Landau pole appears for $g_{B-L}$. The \textit{lower panel}  shows the result of the parameter scan, with color coding depicting the scale at which the scalar potential becomes unstable. It is clear that $m_{Z'},\,m_\chi,\,m_\rho \lesssim 40\,\text{TeV}$, and that if one requires that a Landau pole appears above $\Lambda = 100\,\text{TeV}$ then $g_{B-L}\lesssim2$.}\label{fig:TeV_param}
\end{figure}

In the reduced bound analysis, the low $m_\chi$ regime remains effectively unchanged; we note that a small amount of parameter space has become viable in the vicinity of $m_{Z^\prime} \sim m_Z$, however a slightly wider (\ie $\sim 5$ GeV) resonance cut would have eliminated all of these points. In the intermediate dark matter mass region, a majority of the viable points lie at large values $m_{Z^\prime}$ (\ie $m_{Z^\prime} \gtrsim 4$ TeV), although the number of viable points at $m_{Z^\prime} \lesssim 100$ GeV has also marginally increased. Finally, for $m_\chi > 5$ TeV there exists a large amount of parameter space for both $m_{Z^\prime} \sim \mathcal{O}$(GeV) and $m_{Z^\prime} \gtrsim 4$ TeV (also, as before, there exists a region near the mass of the $Z$ only slightly evading the resonance cuts). An intriguing feature that appears in the high dark matter mass reduced bound analysis ($5\,\text{TeV} \leq m_\chi \leq 6 $ TeV) is the non-existence of viable points for $m_{Z^\prime} \gtrsim 15$ TeV\footnote{We remind the reader that the reduced parameter space scan only searches for viable candidates with gauge couplings lying between the full strength and the reduced bounds.}. This is a feature that arises from the fact that the only constraint on the $Z^\prime$ in this region is from LEP, and those points producing the correct relic abundance are not naturally occurring in the vicinity of this bound (this is shown in \Fig{fig:TeV_param} for the high-mass parameter space scan, to be discussed below).
 
 \begin{table}
\setlength\extrarowheight{5pt}
\begin{center}
\label{table:adaf}
\begin{tabular}{|c|c|c|c|}
\hline\hline 
DM Mass/Surviving Points & $m_\chi \leq $ 900 GeV  & 900 GeV $\leq m_\chi \leq 5 $ TeV  & 5 TeV $\leq m_\chi \leq 6 $ TeV \\ 
 & (Weak Bnd) & (Weak Bnd) & (Weak Bnd) \\
\hline\hline
No Cuts & $100\%$  ($100\%$) & $100\%$  ($100\%$) & $100\%$  ($100\%$) \\
\hline
+ $Z^\prime$ Resonance &  $27.3\%$  ($63.2\%$) & $65.5\%$  ($93.2\%$) &   $84.8\%$   ($81.2\%$) \\
\hline
+ $\rho$ Resonance &      $4.7\%$   ($47.4\%$) & $61.5\%$  ($91.0\%$) &   $83.6\%$   ($79.8\%$) \\
\hline 
+ $Z$ Resonance &         $0\%$     ($8.7\%$) &  $46.4\%$  ($90.2\%$)  &  $78.5\%$   ($79.8\%$)\\
\hline \hline
+ Landau Pole &           $0\%$     ($6.3\%$) &  $46.4\%$  ($80.4\%$) &   $78.5\%$   ($64.6\%$)  \\
\hline
+ Scalar Sector Stability & $0\%$   ($0.8\%$) &  $9.3\%$   ($38.4\%$) &   $34.5\%$   ($24.2\%$) \\
\hline
+ Perturbative &          $0\%$     ($0.8\%$ ) & $6.6\%$   ($36.3\%$ ) &  $29.5\%$   ($24.2\%$ )  \\
\hline \hline
\end{tabular}
\end{center}
\caption{Fraction of sampled points surviving various resonance and RGE cuts. Values in parenthesis correspond to the secondary sampling using the reduced bound analysis described in \Sec{sec:ThC}. The final row contains the fraction of points surviving all cuts.}
\end{table}

Finally, before concluding this section, it is worth commenting on constraints coming from unitarity. The conditions for maintaining unitarity in dark matter models with additional $U(1)$ symmetries have been previously derived in \eg\cite{Duerr:2016tmh,Cox:2017rgn}. These require that the dark matter mass and scalar mass satisfy
\begin{equation}\label{eq:mass_pert}
m_\chi \lesssim \sqrt{\pi}\frac{m_{Z^\prime}}{g_{B-L}} \hspace{0.6cm} {\rm and} \hspace{0.6cm} m_\rho \lesssim \sqrt{\pi}\frac{m_{Z^\prime}}{g_{B-L}} \, ,
\end{equation}
and the gauge and Yukawa couplings to satisfy
 \begin{equation}\label{eq:coupl_pert}
g_{B-L} < \sqrt{4\pi}  \hspace{0.6cm} {\rm and} \hspace{0.6cm} \lambda_{\chi,N} < \sqrt{8\pi}\,.
\end{equation}
While we do not find that enforcing unitarity constrains any of the viable parameter space after applying the aforementioned cuts, we emphasize that such constraints may strongly limit the dark matter mass from above, further constraining this model. More specifically, we find that for values of $m_{Z^\prime}\sim 30$ TeV, typical couplings required to meet the relic density constraints are of order $g_{B-L}\lesssim 2.5$, implying $m_\chi,m_\rho \lesssim 20$ TeV.

In order to make more conclusive statements about the validity of the high-mass region, we perform a specialized scan in which we randomly sample in a linear scale $m_\chi,\,m_{Z'},\, m_\rho,\, m_{N} \in 0.5-100\,\text{TeV}$, and find the $g_{B-L}$ coupling that provides the correct relic abundance (without violating constraints from LEP and the LHC). In \Fig{fig:TeV_param}, we show that off-resonance, the mass of the $Z^\prime$, the dark matter, and the scalar must be $\lesssim 40$ TeV. If one considers resonances, values as high as $m_{Z'} \sim 80\,\text{TeV}$ are attainable, while both $m_\chi,\,m_\rho \lesssim 40\,\text{TeV}$. The color of the points identified in \Fig{fig:TeV_param} also illustrates the scale $\Lambda$ at which either the Landau pole appears, or the scale at which the scalar potential becomes unstable. In order to maintain the validity of this model to moderately large energy scales ($\Lambda \sim 10^4\,\text{TeV})$, the $Z^\prime$ mass must be $\lesssim 20$ TeV. Comparing this number with the maximal $Z^\prime$ mass that can be probed by near-future colliders (\ie $m_{Z'} \lesssim 7\,\text{TeV}$), we see that future HL-LHC constraints are unlikely to provide significant insight into this potentially viable window of parameter space.

%%%%%%%%%%%%%%%%%%%%%%%%%%%%%%%%%%%%%%%%%%%%%%%%%%%%%%%%%%%%%%%%%%%%%%%%%%%%%%%%%%%%%%%%%%
%%%%%%%%%%%%%%%%%%%%%%%%%%%%%%%%%%%%%%%%%%%%%%%%%%%%%%%%%%%%%%%%%%%%%%%%%%%%%%%%%%%%%%%%%%
\section{Non-Minimal Extensions}\label{sec:ext}
%%%%%%%%%%%%%%%%%%%%%%%%%%%%%%%%%%%%%%%%%%%%%%%%%%%%%%%%%%%%%%%%%%%%%%%%%%%%%%%%%%%%%%%%%%
%%%%%%%%%%%%%%%%%%%%%%%%%%%%%%%%%%%%%%%%%%%%%%%%%%%%%%%%%%%%%%%%%%%%%%%%%%%%%%%%%%%%%%%%%%
We have shown that in the minimal model, null collider searches for the $B-L$ gauge boson negate the possibility of thermal dark matter if $m_\chi < 900 \, \text{GeV}$, except for on-resonance regions (\ie $m_\chi \simeq m_{Z'}/2, m_\rho/2$) or near the mass of the SM $Z$ (\ie $m_{Z'} \in \left (m_Z\pm 5 \times \Gamma_Z \right)$). However, such strong constraints only probe the coupling and $Z^\prime$ mass. In the minimal model presented in Section~\ref{sec:Model}, the vev of the scalar is directly related to the mass and coupling of the $Z'$: $m_{Z'} = 2 g_{B-L} v_{B-L}$. However, upon introduction of an additional scalar it is possible to alter this relation, such that the relation between $m_{Z^\prime}$ and $g_{B-L}$ is given by\footnote{We consider only one extra scalar for definiteness, but our conclusions are also applicable to extended models of Majorana fermion dark matter, with any number of  $B-L$ additional charged scalars.}
\begin{eqnarray}
m_{Z'}^2  = g_{B-L}^2 \left(q^2 v_q^2 +4 v_2^2 \right) \, .
\end{eqnarray}
Here, $v_2$ corresponds to the vev of the scalar with charge $-2$ under $B-L$, which has been introduced in order to provide the Majorana mass terms to the sterile neutrinos, and $v_q$ corresponds to the vev of some new scalar. The pertinent question at hand is: how tuned must the ratio $v_2/v_q$ be such that for $m_\chi < 900 \, \text{GeV}$ there are sizable regions of parameter space where the dark matter can be thermally produced without requiring $m_\chi \simeq m_{Z'}/2, m_\rho/2$ or $m_{Z'} \sim m_{Z}$.

 If the charge $q$ is such that the Lagrangian respects the symmetry $\phi_q \to \phi_q^\star$, then a physical Goldstone boson, the Majoron~\cite{Chikashige:1980ui}, will appear in the spectrum. On the other hand, if this symmetry of the scalar potential is absent, then a massive pseudoscalar will form upon symmetry breaking. In order to understand the impact of each scenario we will consider two models. One with $q=1/2$, which therefore includes a physical Goldstone boson in the spectrum, and other with $q=1$ in which a massive pseudoscalar will instead be present.
 
   It is worth stressing that the modification of the $g_{B-L}$ to $m_{Z'}$ relation through the presence of new scalar states is present in a variety of models in the literature, and thus deserves a dedicated study. Note that in both cases we assume that dark matter is one of the three sterile neutrinos (i.e. has $L=1$), so that it couples to $\phi_2$ and thus the annihilation channel $\chi \chi \to N N$ 
 mediated by the scalar and pseudoscalar fields  is not further suppressed by the singlet scalar's mixing angle, $\tan\beta = 2v_2/(q v_q)$, as in models where the dark matter fermion has $L\neq 1$ \cite{DeRomeri:2017oxa}. In this sense, the scenarios analyzed below can be considered as the most favorable ones within the gauged $U(1)_{B-L}$ symmetric paradigm. It is also worth pointing out that the inclusion of additional particle content in a $U(1)_{B-L}$ model leads to an increase in the $\beta$ function for the $B-L$ gauge coupling, and thus generically many extensions may induce a Landau pole at rather low scales.

%%%%%%%%%%%%%%%%%%%%%%%%%%%%%%%%%%%%%%%%%%%%%%%%%%%%%%%%%%%%%%%%%%%%%%%%%%%%%%%%%%%%%%%%%%
\subsection{Additional Sterile Scalar with $q = 1/2$}\label{subsec:q12}
%%%%%%%%%%%%%%%%%%%%%%%%%%%%%%%%%%%%%%%%%%%%%%%%%%%%%%%%%%%%%%%%%%%%%%%%%%%%%%%%%%%%%%%%%%
 Upon the introduction of a new scalar with charge $q=1/2$, one can write the potential as:
\begin{align}
\label{eq:pot}
V =  &- \mu_H^2 |H|^2 + \lambda_H  |H|^4 - \mu_2^2 |\Phi_2|^2 +  \lambda_2  |\Phi_2|^4 -  \mu_q^2 |\Phi_q|^2  + \lambda_q  |\Phi_q|^4\nonumber \\ 
&+\lambda_{H2}  |\Phi_2|^2  |H|^2   +\lambda_{Hq}  |\Phi_q|^2  |H|^2   +\lambda_{2q}  |\Phi_2|^2  |\Phi_q|^2  \,,
\end{align}
where $H$ represents the Higgs doublet. For such charge assignment, the potential has two global symmetries $U(1)_D \times U(1)_{B-L}$, where the first corresponds to $\phi_q \to e^{q_D i\theta} \phi_q$ and the second to $\phi_2 \to e^{2i\theta} \phi_2, \, \phi_q \to e^{qi\theta} \phi_q $~\cite{Rothstein:1992rh}. 

When the scalars develop a vev $\phi_2 = \frac{1}{\sqrt{2}}\left(v_2 +\rho_2 + i\eta_2 \right) $ and $\phi_q = \frac{1}{\sqrt{2}}\left(v_q +\rho_q + i\eta_q \right)$, both symmetries will be spontaneously broken and the minimization of the potential yields two CP-odd massless scalar fields $\eta_2$ and $\eta_q$. The mass matrix for the CP-even scalars will be given by
\begin{eqnarray}
\mathcal{M}^2 \equiv \frac{\partial^2 V}{\partial \phi_i \partial \phi_j} =  \left(
\begin{array}{ccc}
 2 v_H^2 \lambda _H & v_2 v_H \lambda _{{H2}} & v_H v_q \lambda _{{Hq}} \\
 v_2 v_H \lambda _{{H2}} & 2 v_2^2 \lambda _2 & v_2 v_q \lambda _{2 q} \\
 v_H v_q \lambda _{{Hq}} & v_2 v_q \lambda _{2 q} & 2 v_q^2 \lambda _q \\
\end{array}
\right) \,,
\end{eqnarray}
in the $h, \, \rho_2, \, \rho_q$ basis. In order to simplify the phenomenology we will assume that the mixing with the Higgs is negligibly small (\ie $\lambda_{H2} = \lambda_{Hq} = 0$), thus negating various bounds from the LHC. To further simplify the discussion we consider the case where $\lambda _{2 q} = 0$. We note that this last step is not required by any phenomenological reasons, but reduces the number of parameters to be explored. This therefore leads to the following relations: 
\begin{eqnarray}
 m_{\rho_2}^2 = 2 \lambda _2 v_2^2, \qquad  m_{\rho_q}^2 = 2 \lambda _q v_q^2 \, .
\end{eqnarray}
Turning to the phenomenology of the Goldstone boson, after the breaking of both the $U(1)_D$ and  $U(1)_{B-L}$ symmetries, the two massless CP-odd states originally contained in $\phi_2$ and $\phi_q$ mix. One linear combination of these scalars forms the longitudinal mode of the $Z'$ gauge boson and the orthogonal combination renders a massless Goldstone boson. This can be explicitly expressed as
\begin{eqnarray}
 Z'_L =  \frac{q v_q \eta_q + 2 v_2 \eta_2 }{\sqrt{(q v_q)^2+(2v_2)^2}}, \qquad \text{and} \qquad \eta_G =  \frac{2 v_2 \eta_q - q v_q \eta_2 }{\sqrt{(q v_q)^2+(2v_2)^2}} \,,
\end{eqnarray}
where $\eta_G$ is the physical Goldstone boson. Such a relation can be expressed as a rotation of the fields:
\begin{eqnarray}
 \begin{pmatrix}  Z'_L \\ \eta_G \end{pmatrix}  &=&  \begin{pmatrix} \cos \beta & \sin \beta \\ - \sin \beta & \cos \beta \end{pmatrix}  \begin{pmatrix} \eta_q \\ \eta_2 \end{pmatrix} , \qquad \begin{pmatrix} \eta_q \\ \eta_2 \end{pmatrix} =  \begin{pmatrix} \cos \beta & -\sin \beta \\  \sin \beta & \cos \beta \end{pmatrix}   \begin{pmatrix}  Z'_L \\ \eta_G \end{pmatrix},   \qquad \tan \beta \equiv \frac{2 v_2}{q v_q} \, .
\end{eqnarray}
This formulation is convenient as it allows for a linear realization of the scalar fields. Namely, we can write the scalar fields as 
\begin{eqnarray}
\phi_2 = \frac{1}{\sqrt{2}}\left(v_2 +\rho_2  +i\left[Z'_L \sin \beta  + \eta_G  \cos \beta   \right] \right)\,,\, \, \,  \, \, \, \phi_q = \frac{1}{\sqrt{2}}\left(v_q +\rho_q + i\left[ Z'_L \cos \beta  -  \eta_G \sin \beta   \right] \right) \, .
\end{eqnarray}
In the Unitary Gauge, one rotates away the longitudinal modes of the gauge fields, and thus from the scalars one can fully express the interacting Lagrangian for the fermion fields.

Since only $\chi$ and $N$ have Yukawa interactions with the $\phi_2$ scalar, they are the only fields that interact directly with such a boson. The interacting Lagrangian that concerns the fermions and scalars in the model therefore reads
  \begin{eqnarray}
\mathcal{L} &=& \rho_2  \left( \frac{m_\chi}{2 v_2}  \bar{\chi}  \chi + \frac{m_N}{2 v_2}  \bar{N} N \right) - i \eta_G \cos \beta \left( \frac{m_\chi}{2 v_2}  \bar{\chi} \gamma_5  \chi + \frac{m_N}{2 v_2}  \bar{N}\gamma_5  N \right) \, .
 \end{eqnarray}
In the above expression we haven't explicitly written the scalar-scalar and gauge-scalar interactions that will originate from both the scalar potential and the kinetic interaction. We will assume here that $m_{\eta_G} = 0$, as we expect the gravity breaking contributions to be non-perturbative~\cite{Kallosh:1995hi}, and thus negligible. In the case that $m_\eta = 0$, the presence of such a massless particle in the spectrum only leads to a measurable impact in CMB Stage IV experiments~\cite{Abazajian:2016yjj} if it is still in thermal contact with the SM bath for $T < 0.5 \, \text{GeV}$, however such Goldston boson interacts weakly and has momentum suppressed couplings to fermions and thus likely decouples considerably earlier (see \eg\cite{Rothstein:1992rh} for cosmological implications in the case that $m_{\eta_G} \neq 0$).

%%%%%%%%%%%%%%%%%%%%%%%%%%%%%%%%%%%%%%%%%%%%%%%%%%%%%%%%%%%%%%%%%%%%%%%%%%%%%%%%%%%%%%%%%%
\subsection{Additional Sterile Scalar with $q = 1$}\label{subsec:q1}
%%%%%%%%%%%%%%%%%%%%%%%%%%%%%%%%%%%%%%%%%%%%%%%%%%%%%%%%%%%%%%%%%%%%%%%%%%%%%%%%%%%%%%%%%%

Here, we consider the case of an additional sterile scalar with charge $q=1$. After taking $\lambda_{H2} = \lambda_{Hq} = \lambda_{2q} = 0$ for simplicity, the scalar potential reads 
\begin{eqnarray}
\label{eq:pot}
V = - \mu_H^2 |H|^2 + \lambda_H  |H|^4 - \mu_2^2 |\Phi_2|^2 +  \lambda_2  |\Phi_2|^4 -  \mu_1^2 |\phi_1|^2 +  \lambda_1  |\phi_1|^4  - \mu_{21} \left( \phi_2 \phi_{1}^\star \phi_{1}^\star  + \phi_2^\star \phi_{1} \phi_{1}   \right)\,,
\end{eqnarray}
where we take the trilinear coupling $\mu_{21}$ to be real and positive. For such a charge assignment, the mass matrix in the basis $\{h,\rho_2,\rho_q,\eta_2,\eta_q\}$ is given by
\begin{eqnarray}
\mathcal{M}^2 \equiv \frac{\partial^2 V}{\partial \phi_i \partial \phi_j} =   \left(
\begin{array}{ccccc}
 2 v_H^2 \lambda _H & 0 & 0 & 0 & 0 \\
 0 & \frac{\mu _{21} v_1^2}{\sqrt{2} v_2}+2 v_2^2 \lambda _2 & -\sqrt{2} v_1 \mu _{21} & 0 & 0 \\
 0 & -\sqrt{2} v_1 \mu _{21} & 2 v_1^2 \lambda _1 & 0 & 0 \\
 0 & 0 & 0 & \frac{v_1^2 \mu _{21}}{\sqrt{2} v_2} & -\sqrt{2} v_1 \mu _{21} \\
 0 & 0 & 0 & -\sqrt{2} v_1 \mu _{21} & 2 \sqrt{2} v_2 \mu _{21} \\
\end{array}
\right)\,,
\end{eqnarray}
where the minimization condition requires
\begin{eqnarray}
\mu_2^2 = \frac{2 \lambda _2 v_2^3-\sqrt{2} \mu _{21} v_1^2}{2 v_2}\,, \qquad \mu_1^2 = \lambda _1 v_1^2-\sqrt{2} \mu _{21} v_2 \, .
\end{eqnarray}
Upon diagonalization, this leads to the following physical CP-even scalar mixing
\begin{eqnarray}
 \begin{pmatrix} \rho_2' \\  \rho_q' \end{pmatrix}  &=&  \begin{pmatrix} \cos \alpha & \sin \alpha \\ - \sin \alpha & \cos \alpha \end{pmatrix}  \begin{pmatrix} \rho_2 \\  \rho_q \end{pmatrix} , \qquad \tan 2 \alpha = \frac{4 \mu _{21} v_1 v_2}{v_1^2 \left(\mu _{21}-2 \sqrt{2} \lambda _1 v_2\right)+2 \sqrt{2} \lambda _2 v_2^3} \,,
\end{eqnarray}
where the fields $\rho_2' $ and $\rho_q' $ are the CP-even scalars in the interaction basis, and $\rho_2$ and $\rho_q$ are mass eigenstates. In terms of the couplings, the masses of the scalars are
\begin{eqnarray}
& & m_{\rho_2}^2Â \, = Â \, \frac{a-b}{4 v_2} \,, \qquad m_{\rho_1}^2 = \frac{a+b}{4 v_2}\,, \qquad \text{with} Â \\
  a &=& v_1^2 \left(\sqrt{2} \mu _{21}+4 \lambda _1 v_2\right)+4 \lambda _2 v_2^3 \,,\\
b &=& \sqrt{8 \sqrt{2} \mu _{21} v_2 v_1^2 \left(\lambda _2 v_2^2-\lambda _1 v_1^2\right)+16 \left(\lambda _1 v_1^2 v_2-\lambda _2 v_2^3\right){}^2+2 \mu _{21}^2 \left(v_1^2+16 v_2^2\right) v_1^2} \, .
\end{eqnarray}
On the other hand, for the CP-odd states:
\begin{eqnarray} 
m_{Z'_L}^2 = 0 , &\qquad& m_a^2   = \frac{\mu _{21} \left(v_1^2+4 v_2^2\right)}{ \sqrt{2} v_2}	\,,     \\
\begin{pmatrix}  Z'_L \\ a \end{pmatrix}  =  \begin{pmatrix} \cos \beta & \sin \beta \\ - \sin \beta & \cos \beta \end{pmatrix}  \begin{pmatrix} \eta_q \\ \eta_2 \end{pmatrix} , &\qquad& \begin{pmatrix} \eta_q \\ \eta_2 \end{pmatrix} =  \begin{pmatrix} \cos \beta & -\sin \beta \\  \sin \beta & \cos \beta \end{pmatrix}   \begin{pmatrix}  Z'_L \\ a \end{pmatrix}, \hspace{.5cm}   \tan \beta \equiv \frac{2 v_2}{v_1} \, .
\end{eqnarray}
In this model we denote the physical pseudoscalar with $a$ in order to remind the reader that in this case it is massive. Again, as for the model described in \Sec{subsec:q12}, one can express the scalars as
\begin{eqnarray}
\phi_2 &=& \frac{1}{\sqrt{2}}\left(v_2 +\rho_2 \cos \alpha + \rho_q \sin \alpha   +i\left[Z'_L  \sin \beta  + a \cos \beta   \right] \right)\,, \\
  \phi_q &=& \frac{1}{\sqrt{2}}\left(v_q +\rho_q \cos\alpha - \rho_2 \sin \alpha + i\left[ Z'_L \cos \beta - a \sin \beta   \right] \right) \, .
\end{eqnarray}
In the limit $m_a \to 0$, it follows that $\mu_{21} \to 0$, and thus $\lambda_2 \to m_{\rho_2}^2 /(2 v_2^2)$ and $\lambda_1 \to m_{\rho_1}^2 /(2 v_1^2)$.

It is worth mentioning that in order to have a well-defined mass matrix, namely, that the masses of the CP-even scalars are positive, one must require that $\text{det}[\mathcal{M}]_\text{CP-even} > 0$. This results in the following condition for the mass of the pseudoscalar:
\begin{eqnarray}\label{eq:marequirement}
m_a^2 < \frac{\left(v_1^2+4 v_2^2\right) \left(\lambda _1 v_1^2+\sqrt{\lambda _1 \left(\lambda _1 v_1^4+16 \lambda _2 v_2^4\right)}\right)}{4 v_2^2} \,.
\end{eqnarray}

Since the masses of the scalar fields are related to the couplings, we will scan $\lambda_{1,2} \in 10^{-5}-\sqrt{8\pi}$ rather than the masses directly. The mass of the pseudoscalar will be varied from $0.1-30\times 10^{3} $ GeV subject to the consistency requirement given in~\Eq{eq:marequirement}.

%%%%%%%%%%%%%%%%%%%%%%%%%%%%%%%%%%%%%%%%%%%%%%%%%%%%%%%%%%%%%%%%%%%%%%%%%%%%%%%%%%%%%%%%%%
\subsection{Results}\label{subsec:addqres}
%%%%%%%%%%%%%%%%%%%%%%%%%%%%%%%%%%%%%%%%%%%%%%%%%%%%%%%%%%%%%%%%%%%%%%%%%%%%%%%%%%%%%%%%%%
\begin{figure}[h]
\centering
\includegraphics[width=0.99\textwidth]{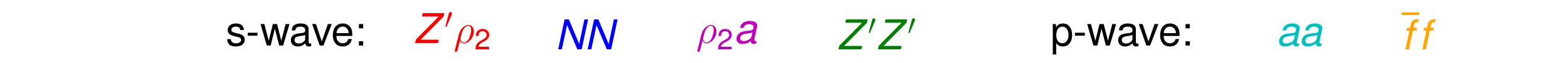}
 \includegraphics[width=0.99\textwidth]{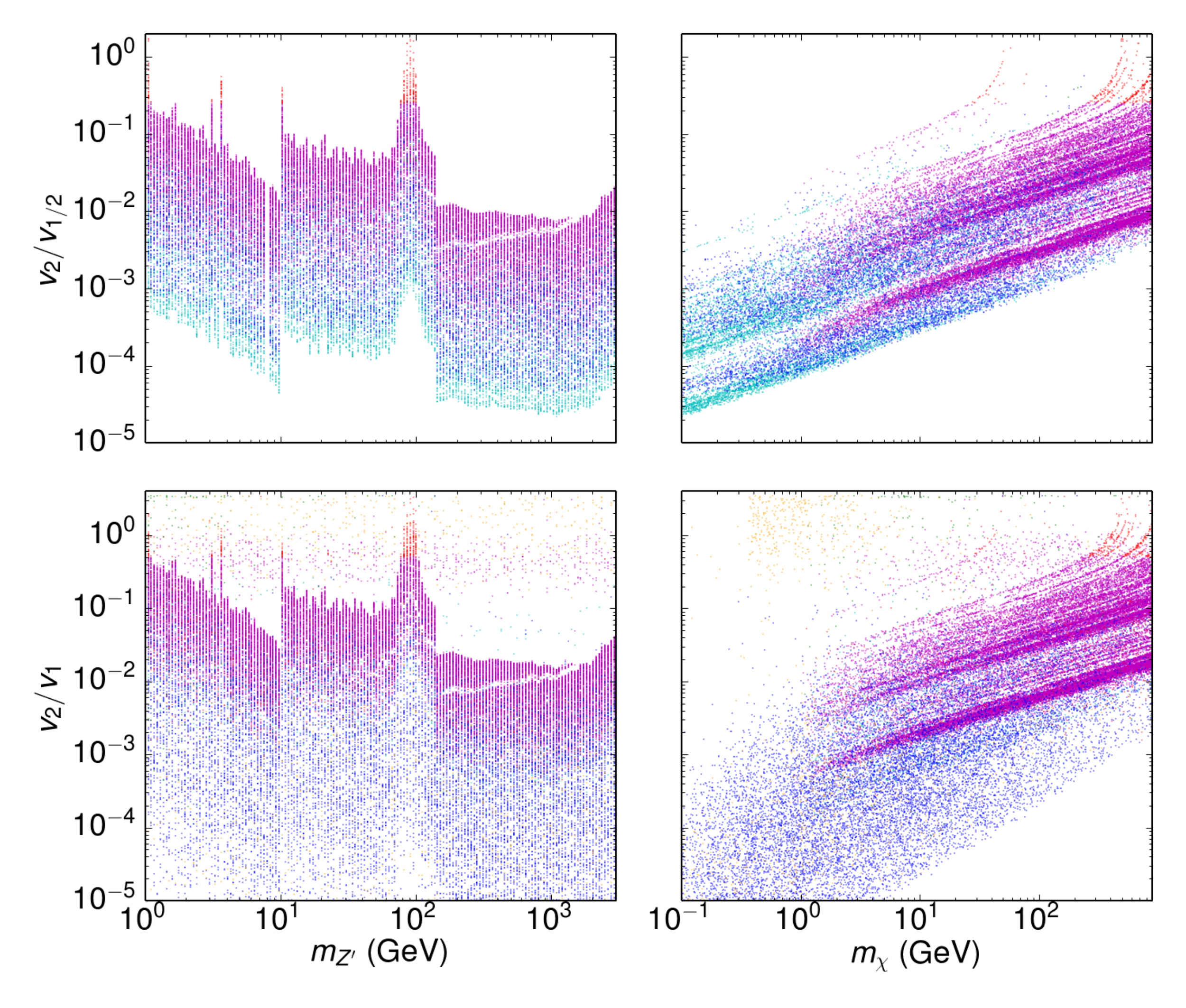}
 \vspace{-0.5cm}
\caption{Viable values of $v_2/v_q$ identified for the non-minimal models described in \Sec{subsec:q12} (top) and \Sec{subsec:q1} (bottom) as a function of the $Z'$ mass. The colors represent the dominant annihilation channel in the contemporary Universe. Resonant points, defined by $ 0.4 < m_\chi/m_{\rho_2}, \,m_\chi/m_{\rho_q}, \, m_\chi/m_a, \, m_\chi/m_{Z'} < 0.6$ have been removed.}\label{fig:tanbeta}
\end{figure}

The inclusion of an additional scalar field changes the dark matter phenomenology due to the appearance of two new annihilation channels. The first, $\chi \chi \to a \, a$ is p-wave, and if $m_a = 0$, it is always open. The second, $\chi \chi  \to a  \, \rho_2$ is s-wave, and if open, dominates over the $\chi \chi \to aÂ \, a$ channel. In addition to the new aforementioned channels, the annihilation $\chi \chi \to NN$ mediated by $a$ and $\rho$ could be enhanced by a factor $v_{B-L}/v_2$, a consequence of the fact that $v_2$ may be lower than in the minimal scenario. In order to draw quantitative conclusions about the required values of $v_2/v_q$ needed to obtain the correct relic abundance without requiring $m_\chi \simeq m_{Z'}/2, \, m_\rho/2$, we performed a parameter scan for both $m_a = 0$ and $m_a > 0$, assuming that the dark matter is relatively light $m_\chi < 900 \, \text{GeV}$ and the $Z'$ mass is $\leq 3 \, \text{TeV}$; we have chosen this region as it is directly testable in terrestrial  experiments. The precise details of the parameter scan are presented in Table~\ref{tab:Parameters_nonminimal}. 

We found that if the scalar masses are such that $ 2m_\chi > m_{\rho_2} + m_a$, then the $\chi \chi \to a  \, \rho_2$ channel will always dominate the annihilation in both the early and the contemporary Universe. If that condition is not met, then the $\chi \chi \to N N$ channel will dominate the annihilation, accompanied by the $\chi \chi \to a \, a$ annihilation, should it be kinematically allowed. Therefore, models with a non-minimal scalar sector allow the dark matter candidate to be produced in the early Universe through the annihilation to the aforementioned channels in a manner similar to what happens in the global $B-L$ scenario (studied at depth in~\cite{Escudero:2016tzx}). The required level of tuning of the vevs, in order not to require $m_\chi \simeq m_{Z'}/2, \, m_\rho/2$, depends on both the $Z'$ mass and the dark matter mass, and typically ranges between $v_2/v_q \in 0.001-0.1$ as shown in \Fig{fig:tanbeta} (note that in \Fig{fig:tanbeta} we have chosen to remove resonant points, defined by $ 0.4 < m_\chi/m_{\rho_2}, \,m_\chi/m_{\rho_q}, \, m_\chi/m_a, \, m_\chi/m_{Z'} < 0.6$). 

As shown in~\cite{Escudero:2016tzx}, for the case $m_a = 0$ unless $m_\chi \gtrsim m_N > 100 \, \text{GeV} $ the final state of s-wave annihilations will lead to invisible particles. The reason is that the main decay of light sterile neutrinos ($m_N \lesssim m_W$) will be $ N \to \nu \, a$ which does not render gamma-rays. Similarly, the decay of the $\rho_2$ scalar will be $\rho_2 \to a \, a$, and is thus also invisible. If the pseudoscalar is not massless the phenomenology may differ. In this case, the massive pseudoscalar will always decay to light active neutrinos unless $m_N <  m_a /2$. Therefore, if the annihilation is dominated by $\chi \chi \to N N$ the sterile neutrinos will decay to SM particles, and $m_\chi \lesssim 100 \, \text{GeV}$ are excluded from dwarf galaxies observations~\cite{Campos:2017odj,Batell:2017rol,Folgado:2018qlv}. Dark matter masses $m_\chi \lesssim 100$ GeV are also excluded from gamma-ray observations in the case that the annihilation is dominated by $\chi \chi \to a  \, \rho_2$, provided that the sterile neutrinos are light enough, \ie $m_N < m_\rho/2,\, m_a/2  $.
 In the case that $\chi \chi \to a  \, \rho_2$ dominates and $m_N> m_\rho/2,\, m_a/2$, then the final state will be entirely composed of light active neutrinos. Current bounds from neutrino telescopes on such annihilation channels are about 3 orders of magnitude above $\left< \sigma v\right> = 3 \times 10^{-26} \, \text{cm}^3/\text{s}$~\cite{Aartsen:2017ulx,Albert:2016emp} and thus this portion of parameter space will remain elusive.

\begin{table}[h]
\begin{center}
\begin{tabular}{ccc}
\hline\hline
  Parameter       &		 Description			 & 		Range     	       \\
\hline
  $m_\chi$         & Dark matter mass 	 & 	  $[0.1-900]$ GeV   \\
    $m_{N_1}=m_{N_2}$         & Sterile Neutrino mass  			 & 	  $[0.1-3\times10^4]$ GeV      \\
            $m_{\rho_2}, \, m_{\rho_q}  \, $&  CP-even masses &   $[0.1-3\times10^4]$ GeV   	  \\
    $m_{a}$&  Pseudoscalar mass &   $[0.1-3\times10^4]$ GeV   	  \\
    $m_{Z'}$&  $Z'$ mass &   $[1-3\times10^3]$ GeV   	  \\
    $g_{B-L}$& $B-L$ coupling &   $g_{B-L} = g_{B-L}^{2\sigma}(m_{Z'}) $   	  \\
  \hline \hline
\end{tabular}
\end{center}
\caption{Ranges of the parameters explored for the non-minimal models. Values for $m_{Z^\prime}$ were sampled using $154$ logarithmically spaced points. The value of the $g_{B-L}$ coupling was not taken as a free parameter, but instead fixed to the collider $2\sigma$ exclusion value for each $m_{Z'}$ value. For each point random values of $m_\chi, \, m_{N_1}=m_{N_2}, \, m_{\rho_2}, m_{\rho_q} , \, m_a$ were drawn from the range detailed in the table above (assuming a log-flat distribution, except for $m_a$ in the $q=1/2$ case where $m_a = 0$), and $\tan \beta$ or equivalently $v_2$ was set to be consistent with the measured relic abundance at $2\sigma$, \ie $\Omega_{DM} h^2 = 0.120 \pm 0.003$~\cite{Ade:2015xua}. We further require all coupling to be $< \sqrt{8\pi}$ in order for the model to remain perturbative. The total number of points in parameter space that meet such requirements are $1.54 \times 10^5$ for both the $q=1/2$ and $q=1$ cases. }\label{tab:Parameters_nonminimal}
\end{table}

%%%%%%%%%%%%%%%%%%%%%%%%%%%%%%%%%%%%%%%%%%%%%%%%%%%%%%%%%%%%%%%%%%%%%%%%%%%%%%%%%%%%%%%%%%
\section{Conclusions}\label{sec:Conclusions}
%%%%%%%%%%%%%%%%%%%%%%%%%%%%%%%%%%%%%%%%%%%%%%%%%%%%%%%%%%%%%%%%%%%%%%%%%%%%%%%%%%%%%%%%%%

In this work we have presented a thorough investigation of the simplest extension of the Standard Model that generates active neutrino masses and contains a viable dark matter candidate within the context of a gauged $U(1)_{B-L}$ symmetry. A number of groups have addressed aspects of similar models, \eg by considering a Dirac dark matter candidate (for which direct dark matter detection experiments are strongly constraining)~\cite{Alves:2015mua,DeRomeri:2017oxa,FileviezPerez:2018toq,Han:2018zcn} or a simplified subset of parameter space~\cite{Kaneta:2016vkq,Okada:2016gsh,Klasen:2016qux,Okada:2018ktp,Okada:2010wd,Okada:2012sg}, however until now a comprehensive understanding of the viability of this model has been absent from the literature. Specifically, we take one of the right-handed neutrinos required to achieve anomaly cancellation to be a Majorana particle and stable under a new symmetry, thus providing a dark matter candidate with suppressed direct detection constraints. 
Although this additional stabilizing symmetry may seem {\it ad hoc}, we have explored the minimal model because it captures
the main features of Majorana  dark matter phenomenology within gauged $U(1)_{B-L}$.
We investigate the unavoidable limits from various colliders on the existence of the new gauge boson (which are fairly independent of the Dirac or Majorana nature of the dark matter), and consider deeper theoretical concerns such as the appearance of a Landau pole or instability of the scalar sector arising at energies as low as the TeV scale.

After performing a full parameter space scan, our results reveal a number of important points: \emph{(i)} dark matter with mass $\lesssim 900$ GeV must annihilate on-resonance to remain viable, \emph{(ii)} high energy dilepton searches rule out $150$ GeV $< m_{Z^\prime} < 3$ TeV (even when annihilations proceed on-resonance), and \emph{(iii)} the theoretical considerations, such as annihilations that proceed on-resonance, the appearance of a Landau pole, or the instability of the scalar sector  at very low energies strongly disfavors models in which $m_\chi, m_{Z^\prime} \lesssim 5$ TeV.  While unitarity and perturbativity constraints require $m_\chi, m_{Z^\prime} \lesssim 40$ TeV, we have also shown that avoiding low-energy Landau poles and instability in the high-mass region likely require $m_{Z^\prime} \lesssim 20$ TeV. Unfortunately, constraints from the HL-LHC cannot close this window (we estimate the maximal sensitivity to be approximately $m_{Z^\prime} \lesssim 7$ TeV), implying the small remaining viable and theoretically well-motivated parameter space is largely untestable. Thus, we believe it is safe to say that the minimal gauged $U(1)_{B-L}$ extension of the Standard Model with right-handed neutrino dark matter is under siege, and should at this point be considered either a disfavored or untestable extension of the Standard Model. 

The stringent constraints on the minimal model raise the question of whether minor modifications could relieve existing tension with experimental bounds. Perhaps the simplest, however entirely representative, extension for which this can be achieved involves introducing a new scalar particle; this modification allows one to decouple the relation between the vev of the scalar and the ratio of $m_{Z^\prime}$ to $g_{B-L}$. We have considered two such extensions, one which gives rise to a physical Goldstone boson and the other which gives rise to a massive pseudoscalar. We investigate the viable parameter space in these models, and show that typically the tuning between the ratio of the vevs of the $B-L$ charged scalars must be $\sim 10^{-2}$. The addition of more complicated particle content to resolve the aforementioned difficulties inherently increases the instability of the $\beta$ functions, potentially yielding problems at rather low scales. 

In summary, constraints on the minimal gauged $U(1)_{B-L}$ extension of the Standard Model with thermal right-handed neutrino dark matter significantly reduce its current appeal. Minor modifications via the introduction of a new scalar can revive the model, however only at the cost of a potential fine-tuning of the vevs of the scalar fields. Given the theoretical motivations for this model, it is conceivable that non-thermal dark matter production mechanisms could revitalize interest in right-handed neutrino dark matter within a gauged $U(1)_{B-L}$ extension.

\section*{Acknowledgements} 
This work has been partially supported by the European Union's Horizon 2020 research and innovation programme under the Marie Sklodowska-Curie grant agreements No 674896 and 690575, 
by the Spanish MINECO grants FPA2014-57816-P,  FPA2017-85985-P and  SEV-2014-0398,
and by Generalitat Valenciana grant PROMETEO/2014/050.
ME is supported by Spanish Grant FPU13/03111 of MECD. 
SJW is supported by the European Union's Horizon 2020 research and innovation program under the Marie Sk\l{}odowska-Curie grant agreement No. 674896.

  %%%%%%%%%%%%%%%%%%%%%%%%%%%%%%%%%%%%%%%%%%%%%%%%%%%%%%%%%%%%%%%%%%%%%%%%%%%%%%%%%%%%%%%%%%
\appendix   
\section{Appendix: Cross Sections}\label{subsec:Xsecs}
%%%%%%%%%%%%%%%%%%%%%%%%%%%%%%%%%%%%%%%%%%%%%%%%%%%%%%%%%%%%%%%%%%%%%%%%%%%%%%%%%%%%%%%%%%
For completeness we list the various annihilation cross sections, in expansions of velocity $v$, for annihilations to SM fermions, right-handed neutrinos, two $Z^\prime$s, two scalars $\rho$, and one $Z^\prime$ and one scalar $\rho$:

\begin{align}
\sigma v_{\chi\chi\to \bar{f}f} = N_c \frac{g_{B-L}^4 m_\chi^2}{6 \pi  \left(m_{Z'}^2-4 m_\chi^2\right)^2} \, v^2 + \mathcal{O}(v^4) \, ,
\end{align}

\begin{align}
\sigma v_{\chi\chi\to NN} = \frac{g_{B-L}^4 m_{N}^2 \sqrt{m_\chi^2-m_{N}^2}}{4 \pi  m_\chi \left(m_{Z'}^2-4 m_\chi^2\right)^2} + \mathcal{O}(v^2) \,,
\end{align}

\begin{align}
 	 \sigma v_{\chi\chi\to Z'Z'} = \frac{g_{B-L}^4}{ 4 \pi  m_\chi \left(m_{Z'}^2-2 m_\chi^2\right)^2}  \left[m_\chi^2-m_{Z'}^2\right]^{3/2} + \mathcal{O}(v^2) \, ,
\end{align}

\begin{align}
 	 \sigma v_{\chi \chi \to Z' \rho} = \frac{g_{B-L}^4 }{64 \pi  m_\chi^4 m_{Z'}^4}\left[m_{Z'}^4 +\left(m_\rho^2-4 m_\chi^2\right)^2-2 m_{Z'}^2 \left(4 m_\chi^2+m_\rho^2\right)\right]^{3/2}+ \mathcal{O}(v^2)
\end{align}

\begin{align}
	 \sigma v_{\chi \chi \to \rho \rho} &=\frac{g_{B-L}^4 m_\chi \sqrt{m_\chi^2-m_\rho^2} \left(8 m_\chi^4-8 m_\chi^2 m_\rho^2+3 m_\rho^4\right) }{24 \pi  m_{Z'}^4 \left(m_\rho^2-4 m_\chi^2\right)^2 \left(m_\rho^2-2 m_\chi^2\right)^4} \, v^2\\
	 &\times \bigg[288 m_\chi^8-352 m_\chi^6 m_\rho^2+200 m_\chi^4 m_\rho^4-64 m_\chi^2 m_\rho^6+9 m_\rho^8\bigg]  + \mathcal{O}(v^4) \, . \nonumber
\end{align}

We also include the relevant direct detection cross sections arising from the interaction of the $Z^\prime$:
\begin{equation}
\sigma^\text{SI, Z'}_{\chi N} = v_\perp^2  \frac{g_{B-L}^4}{\pi} \frac{\mu_{\chi N}^2}{m_{Z'}^4}\,,
\end{equation}
where $m_N$ is the nucleon mass and $\mu_{\chi N}$ is the $\chi-N$ reduced mass. The scattering cross section as mediated by Higgs/$\rho$:
\begin{equation}
\sigma^\text{SI, H}_{\chi N} =  \sin^2 2\theta \,
\frac{f_N^2}{4 \pi }  \,  \frac{\mu_{\chi N}^2 m_\chi^2 m_N^2 }{v_H^2
v_{B-L}^2} \left(\frac{1}{m_\rho^2} - \frac{1}{m_H^2} \right)^2 \, ,
\end{equation}
where $f_N\simeq 0.3$, and $\theta$ represents the Higgs-$\phi$ mixing angle.

\bibliography{biblio} 
 
 \end{document}